\documentstyle[epsfig]{article}

\renewcommand{\thesection}{\arabic{section}}

\newcommand{\gtwid}{\mathrel{\raise.3ex%
\hbox{$>$\kern-.75em\lower1ex\hbox{$\sim$}}}}
\newcommand{\ltwid}{\mathrel{\raise.3ex%
\hbox{$<$\kern-.75em\lower1ex\hbox{$\sim$}}}}

\newcommand{\rhomega}{\rho_{\omega}}
\newcommand{\rhalpha}{\rho_{\alpha}}

\begin{document}


\title{Observation of the Final Boundary Condition:\\
Extragalactic Background Radiation\\ and the Time Symmetry of the
Universe\thanks{UCSBTH-95-8; gr-qc/9508004}}

\author{David A. Craig\thanks{E-mail: dcraig@hawking.phys.ualberta.ca}\\
{\it Department of Physics, University of California-Santa Barbara,}\\
{\it Santa Barbara, California 93106-9530, USA\thanks{Present Address:
Theoretical Physics Institute, University of Alberta, 
Edmonton, Alberta T6G 2J1}}}

\maketitle

\begin{abstract}

This paper examines an observable consequence for the diffuse extragalactic
background radiation (EGBR) of the hypothesis that if closed, our universe
possesses time symmetric boundary conditions.  By reason of theoretical
and observational clarity, attention
is focused on optical wavelengths.  The universe is modeled as closed
Friedmann-Robertson-Walker.  It is shown that, over a wide range of
frequencies, electromagnetic radiation can propagate largely unabsorbed 
from the present epoch into the recollapsing phase, confirming and
demonstrating the generality of results of Davies and Twamley \cite{DT}.  
As a consequence, time symmetric boundary conditions imply that the optical 
EGBR is at least twice that due to the galaxies on our past light cone, and 
possibly considerably more.  It is therefore possible to test 
 {\it experimentally} the notion that if our universe is closed, it may be 
in a certain sense time symmetric.  The lower bound on the ``excess" EGBR 
in a time symmetric universe is consistent with present observations. 
Nevertheless, better observations and modeling
may soon rule it out entirely.  In addition, many physical
complications arise in attempting to reconcile a transparent future
light cone with time symmetric boundary conditions, thereby providing
further arguments against the possibility that our universe is time
symmetric.  This is therefore a demonstration by example that
physics today can be sensitive to the presence of a boundary condition 
in the arbitrarily distant future.

\end{abstract}

\section{Introduction}
\label{sec:egbrintroduction}
\setcounter{equation}{0}

Quantum cosmology studies the relation between the observed universe and
its boundary conditions in the hope that a natural {\it theory} of the 
boundary condition might emerge (see \cite{Halliwell} for an outstanding
review of this enterprise.)  Assessment of a particular theory requires an 
understanding of its implications for the present day.  To that end, this 
paper elaborates on work of Gell-Mann and Hartle \cite{TSA} and Davies and 
Twamley \cite{DT} by examining the observable consequences for the diffuse
extragalactic background radiation (EGBR) of one possible class of boundary 
conditions, those that are imposed time symmetrically at the beginning 
and end of a closed universe \cite{TSA}, and sketches some of the 
considerable difficulties in rendering this kind of model credible.
Assuming such difficulties do not vitiate the consistency of
time symmetric boundary conditions as a description of our universe, 
the principal conclusion is that these boundary conditions imply that the 
bath of diffuse optical radiation from extragalactic sources be at least 
twice that due only to the galaxies to our past, and possibly much more.
In this sense, observations of the EGBR are observations of the final
boundary condition.
This conclusion will be seen to follow (section \ref{sec:limit}) because 
radiation 
from the present epoch can propagate largely unabsorbed until the universe 
begins to recollapse \linebreak 
 (\cite{DT}, and section \ref{sec:opacity}), even if the 
lifetime of the universe is
very great. By time symmetry, light correlated with the thermodynamically
reversed galaxies of the recollapsing phase must exist at the present 
epoch.  The {\it minimal} predicted ``excess" EGBR in a universe with 
time symmetric boundary conditions turns out to be 
consistent with present observations (section \ref{sec:observations}),
but improved observations and modeling of galactic evolution will
soon constrain this minimal prediction very tightly.  
In addition, many physical
complications with the ansatz that time symmetric boundary conditions
provide a reasonable and consistent description of the observed universe 
will become apparent.  Thus this work may be viewed as outlining some 
reasons why even if very long-lived, our universe is probably 
not time symmetric.

The plan of the paper is as follows.   Section \ref{sec:motivations}\ 
discusses a model universe that will define the terms of the investigation.  
Section \ref{sec:tsbc}\ provides some perspective on doing physics with 
boundary conditions at two times with an eye toward section 
\ref{sec:difficulties}, where some aspects of the
reasonableness of two time boundary conditions not immediately related to
the extragalactic background radiation are discussed.  Section
\ref{sec:opacity}\ generalizes and confirms
work of Davies and Twamley \cite{DT} in showing that for processes of
practical interest, our future light cone (FLC) is transparent all the 
way to the recollapsing era over a wide range of frequencies, even if the 
universe is arbitrarily long-lived.  Section \ref{sec:limit}\ explains 
why this fact 
implies a contribution to the optical extragalactic background radiation
in a universe with time symmetric boundary conditions in excess of 
that expected without time symmetry.  In the course of this explanation,
some rather serious difficulties will emerge in the attempt to reconcile 
time symmetric boundary conditions, and a transparent future light cone, 
into a consistent model of the universe which resembles the one in which 
we live.  Section \ref{sec:observations}\ compares the predictions of
section \ref{sec:limit}\ for the optical EGBR to models of the 
extragalactic background light and observations of it.  
Section \ref{sec:summation}\ is reserved for summation and conclusions.

\subsection{Motivations and A Model}
\label{sec:motivations}

The possibility that the universe may be time symmetric has been raised 
by a number of authors 
 \cite{Gold,Wheeler1,Wheeler2,Hawking1,Zeh1,Zeh2,Zeh3,KZ}.  
Of course, what is meant is not {\it exact} time symmetry, in the sense that 
a long time from now there will be another Earth where everything happens 
backwards.  Rather, the idea is that the various observed ``arrows of time" 
are directly correlated with the expansion of the universe, consequently 
reversing themselves during a recontracting phase if the universe is closed.  
Of central interest is the thermodynamic arrow of entropy increase, 
from which other time arrows, such as the psychological arrow of 
perceived time or the arrow defined by the retardation of radiation, 
are thought to flow \cite[are some reviews]{PTA,Zeh2,HPM}.
However, the mere reversal of the universal expansion is insufficient to
reverse the direction in which entropy increases 
\cite{Schulman1,PTA,Wheeler1,Wheeler2,Penrose1,Page1,Hawking2,HLL}.  
In order to construct a quantum physics for matter in a 
recollapsing universe in which the thermodynamic arrow naturally reverses 
itself, it appears necessary to employ something like the time neutral 
generalization of quantum mechanics \cite{Cocke,ABL,griffiths84,Unruh,TSA} 
in which boundary conditions are imposed near both the
big bang {\it and} the big crunch.  These boundary conditions take the form 
of ``initial" and ``final" density operators\footnote{I retain this
terminology even though these operators may not be of trace class, for
example in the familiar case where the final ``boundary condition" is
merely the identity operator on an infinite dimensional Hilbert space.}
which, when CPT-reverses of one another, define what is meant here by a 
time symmetric universe.\footnote{In other words, the effective 
decoherence functional for matter in a time symmetric universe 
is a canonical decoherence functional $d(h,h') = 
 {\rm tr}[\rhomega h^{\dagger}\rhalpha h']/({\rm tr}\rhalpha\rhomega)$,
with $\rhomega = ({\cal CPT})^{-1}\rhalpha ({\cal CPT})$ \cite{TSA}.}
In such a model the collection of quantum mechanical
histories is time symmetric in the sense that each history in a
decohering set ({\it i.e.\ }a set of histories in which relative
probabilities may be consistently assigned) occurs with the same 
probability as its CPT-reverse \cite{TSA,Page2}.  (As CP violation 
is small, there is for many purposes no difference between CPT- and 
T-symmetry with a T invariant Hamiltonian.)  I do not describe these ideas 
in further detail because very little of the {\it formalism} of generalized 
quantum theory will be directly applied in this paper, 
but it is worth mentioning that in order for the resulting time
symmetric quantum theory to have non-trivial predictions, 
the initial and final density operators must not represent pure 
states.\footnote{As the first in an occasional series of
comments directed to those familiar with the ideas of generalized quantum 
theory, this is because if the initial and final density operators are 
pure, at most two histories can simultaneously be assigned probabilities 
\cite{TSA,dhgqm,craig95}, {\it i.e.\ }the maximum number of histories 
in any weakly decohering set is two!  
Complete information about the history of the universe must be encoded 
in the boundary conditions.  Details of the formalism of generalized 
quantum theory may be found in, {\it e.g.,} \cite{LesH}.  Generalized 
quantum mechanics with boundary conditions at two times is discussed 
more extensively in \cite{TSA} and \cite{dhgqm,craig95}.}  
Interesting theories therefore have boundary conditions 
which are quantum statistical ensembles.

The interest in applying this class of quantum theories (namely, 
theories with CPT-related boundary conditions) in cosmology
lies in the idea that the manifest arrow of time we observe is an
emergent property of the universe, and not built directly into its
structure by asymmetric dynamical laws or an asymmetric choice of
boundary conditions.  Dynamical laws are believed to be
CPT-symmetric, so an asymmetric choice of boundary conditions is usually
cited as the explanation for the existence of a definite arrow of 
time which flows in the same direction throughout the observable part of
spacetime \cite{PTA,Zeh2,HPM}.  However, it is worth investigating whether 
this assumption is {\it required} of us by making the alternate, apparently 
natural ansatz that the boundary conditions on a closed universe are (in a 
relevant sense) equivalent at the beginning and end of time, instead of the 
more usual assumption that the initial condition is somehow special and the 
final condition one of ``indifference," {\it i.e.\ }determined entirely by 
the past.  Another point of view is that, as noted in \cite{TSA}, these 
alternative choices are in some sense opposite extremes.  It is therefore 
of interest to determine whether they are distinguishable on experimental 
grounds, employing time symmetric boundary conditions as a laboratory for 
testing the sensitivity of physical predictions to the presence of a final 
boundary condition.

For the benefit of those eager to proceed to the definite physical 
predictions of sections \ref{sec:opacity}\ and \ref{sec:egbr}, I now 
specify a model in which they might be expected to arise.  
(The cautiousness of this statement is explained in the sequel.)
Sections \ref{sec:tsbc}, \ref{sec:amplifications}, 
 and \ref{sec:difficulties}\ elaborate on the physics expected 
in a universe with
two-time, time symmetric boundary conditions; here I merely summarize
what is required from those sections for a complete statement of the 
assumptions.

   
The model of the universe considered here consists in:
\begin{itemize}
\item   A fixed closed, homogeneous and isotropic background spacetime,
 {\it viz.} a $k=+1$ Friedmann-Robertson-Walker (FRW) universe.  
The evolution
of the scale factor is determined from Einstein's equations by the
averaged matter content of the universe.  (Inhomogeneities in the matter 
content can be treated as additional matter fields on this background.)

\item Boundary conditions imposed on the matter content 
through a canonical decoherence functional $d_{\alpha\omega}$ \cite{dhgqm}
with CPT-related density operators $\rho_{\alpha}, \rho_{\omega}$ 
that describe the
state of matter at some small fiducial scale factor, near what would in 
the absence of quantum gravitational effects be the big bang and big 
crunch, but outside of the quantum gravity regime.  The matter state 
described by one of
these density operators reflects the presumed state of the early universe, 
namely matter fields in apparent thermal equilibrium at the temperature
appropriate to the fiducial scale factor and the total amount of matter
in the universe.  Spatial fluctuations should be consistent with present
day large scale structure, say being approximately scale invariant and 
leading to an amplitude of order $10^{-5}$ at decoupling in order to be 
consistent with recent COBE results \cite{COBE}.  Possible further 
conditions on $\rho_{\alpha}, \rho_{\omega}$ are discussed at the end 
of this subsection.
\end{itemize}


The essence that a choice of boundary conditions with these apparently 
natural characterisics intends to capture is that of a universe in which 
the cosmological principle holds, which is smooth (and in apparent thermal 
equilibrium) whenever it is small, and which displays more or less familiar 
behaviour when larger.  Most of the conclusions of this investigation 
really only 
rely on these general properties, but for the sake of definiteness a 
fairly specific model which has the right general physical characteristics 
is offered.
However, as will be repeatedly emphasized in the sequel, in models with 
boundary conditions at two times, not only does the past have implications 
for the future, but the future has implications for the past.  Therefore, 
in any attempt to model the {\it observed} universe with time symmetric
boundary conditions, we need to make sure it is possible to integrate them 
into a self-consistent picture of the universe as we see it today.
This means in particular that we must be prepared
for the possibility that the early universe in a time symmetric universe
may have properties different from those expected in a universe with an 
initial condition only.  The model boundary conditions sketched above are 
not intended to be so restrictive as to rule out such differences, and 
hence must be taken with a grain of salt or two.  The issue is then 
whether or not these properties are consistent with observation.  
Indeed, the prediction of an ``excess" optical EGBR (to be discussed 
in section \ref{sec:egbr}) in a universe with time symmetric boundary 
conditions is precisely of this character.  (Of course, the most 
extreme possibility is that a universe
burdened with these boundary conditions would look {\it nothing at all}
like the universe in which we actually live.)  Variations on this theme 
will recur frequently in the following sections.  

With these boundary conditions \cite{TSA}, the time neutral 
generalized quantum mechanics of Gell-Mann and Hartle defines 
an effective quantum theory for
matter in the universe which may be imagined to arise from some other, 
more fundamental, quantum cosmological theory of the boundary conditions.
(Presumably, the fundamental quantum cosmological  decoherence functional 
incorporates the 
quantum mechanics of the gravitational field as well, 
a complication that is not addressed in this paper.)

In this connection it is perhaps worth mentioning that it was once 
claimed \cite{Hawking1} that the no-boundary proposal for the initial 
condition\cite[\cite{Halliwell} is an excellent review]{HH} implied just
this sort of effective theory in that it appeared to require that the 
universe be smooth whenever it was small.  Thus a fundamental theory of 
an initial condition {\it only} apparently could be decribed by an
effective theory with {\it two-time} boundary conditions.  However, this 
claim has since been recanted \cite{Hawking2} due to a mathematical
oversight.  More generally, the no-boundary wave function does not 
appear to be a good candidate for a boundary condition imposed time 
symmetrically at both ends of a closed universe because it is a pure 
state \cite{TSA}, which as noted above yield quantum theories in which
essentially all physical information is encoded in the boundary
conditions alone.  For contrast, see \cite{KZ}, in which it is asserted 
that the only sensible way to interpret the Wheeler-DeWitt equation
necessitates a boundary condition requiring the universe to be 
smooth whenever it is small.

In order to allow definite predictions to be made, the final key assumption 
of the model is that, for a suitable class      
of physically interesting coarse grainings ({\it e.g.\ }coarse grainings
defining the domain of classical experience, or local quantum mechanics
experiments), the probabilities for such coarse grained physical histories 
unfold near either boundary condition (relative to the total lifetime of the
universe) in a fashion insensitive to the presence of the boundary
condition at the other end, {\it i.e.\ }as if the other boundary condition
were merely the identity.  As discussed in section \ref{sec:tsbc}, 
simple stochastic models \cite{Cocke,TSA,Schulman2,Schulman3,Schulman4} 
suggest that this holds for
processes for which the ``relaxation time"      
of the process to equilibrium is short compared to the total time between 
the imposition of the boundary conditions (see section \ref{sec:tsbc}), 
which expectation is 
rigorously supported in the case of Markov processes \cite{Schulman5}.  
Therefore, for the sake of brevity this predictive ansatz shall often be 
referred to as the ``Relaxation Time Hypothesis" (RTH).  A more careful 
statement of the RTH would identify the classes of coarse grained histories
(presumably at least those decribing short relaxation time processes at
times sufficiently close to, for instance, the big bang) 
for which conditional probabilities (when defined) are, in a universe with 
boundary conditions $\rho_{\alpha}$, $\rho_{\omega}$, supposed to be close 
to those of a universe with $\rho_{\omega}=1$.\footnote{Some subtleties 
regarding 
equivalences between boundary conditions at one and two times are being 
concealed here, for which see \cite{craig95}.}      
It would also define ``close" and ``short relaxation time process" more 
precisely.  (Thus, a rigorous statement of what is meant by the RTH
requires a definite mathematical model.  Because the concerns of this
investigation  
encompass a variety of complicated physical processes, with the entire 
universe considered as a single physical system, I do not attempt that 
here.  In specific cases the intuitive content of the ansatz ought 
to be clear enough.)  In order to exploit the RTH to its fullest, I shall, 
when convenient, assume also that the universe is close to the critical
density, so that its lifetime is very long.  This plausibly realistic
assumption maximizes the possibility that a model of the kind given
above accurately describes our universe.

With these assumptions, this model universe might be expected to closely
resemble the universe as it is observed today {\it if} most familiar and
important physical processes are examples of such ``short relaxation
time processes.''  In particular, under the assumption that they are,
predictions we expect of a single initial condition $\rho_{\alpha}$ only 
({\it i.e.\ }$\rho_{\omega}= 1$) can (by the RTH) 
be assumed to hold near the initial condition in the model with 
CPT-related initial and final conditions.  Such predictions should 
include those regarding
inflation, relative particle abundances, and the formation of large (and
small) scale structure.  Because of the CPT-related boundary condition
at the big crunch, a qualitatively similar state of affairs is then 
expected in the recollapsing era, but CPT reversed.  As the
thermodynamic arrow is caused fundamentally by gravitational collapse
driving initially smooth matter away from 
equilibrium \cite[for example]{Penrose1,Zeh2}, the arrow of 
entropy increase near the big crunch will run in the opposite direction 
to that near the big bang.\footnote{Some discussion of the state of the 
universe when it is large, which somehow must interpolate between these 
opposed thermodynamic arrows, is provided in sections \ref{sec:tsbc},
\ref{sec:amplifications}, and 
\ref{sec:difficulties}.}  Observers on planets in the recollapsing phase 
will find their situation indistinguishable from our own, with all time
arrows aligned in the direction of increasing volume of the universe.
It is this interesting (if unconventional) state of affairs which leads
to the conclusion that observations of the EGBR can reveal the presence
of a final boundary condition that is CPT-related to the initial, even 
if the lifetime of the universe is very great.  How this comes about is 
the topic of section \ref{sec:egbr}.

Before proceeding, some points made in the preceeding paragraph require 
qualification.  (Readers whose only interest is the extragalactic 
background 
radiation should procede directly to section \ref{sec:opacity}.  This 
discussion, and that of the next subsection, are positioned here for 
the sake of unity and perspective.)  First, the ``relaxation times" for 
many important physical processes in a recollapsing universe 
 {\it do not} appear to be short compared to the lifetime of the universe, 
even if that lifetime is arbitrarily long.  The result described in section
\ref{sec:opacity}, that light from the present epoch can propagate 
unabsorbed into the recollapsing phase, is an excellent example of this 
\cite{DT,TSA}.  Some physical difficulties with the consistency of time 
symmetry to which this gives rise will become apparent in section 
\ref{sec:amplifications}.  Further examples of such physical
difficulties relating to issues such as gravitational collapse, the 
consequent 
emergence of a familiar thermodynamic arrow, and baryon decay in a universe 
with time symmetric boundary conditions are discussed briefly in section
\ref{sec:difficulties}.    
The self-consistency of such models is thus in doubt, with these kinds
of complications constituting arguments against the possibility that our 
universe possesses time symmetric boundary conditions.  That is, it appears 
that time symmetry is {\it not} consistent with the central predictive
ansatz that physics near either boundary condition is practically 
insensitive 
to the presence of the other, and it appears likely that a universe with
time symmetric, low (matter) entropy boundary conditions would look
 {\it nothing like} the universe in which we actually live.  
Nevertheless, the strategy of sections \ref{sec:opacity}\ 
and \ref{sec:egbr}\ 
is to assume that the time symmetric picture is consistent with the gross 
characteristics of the observed universe and see what it predicts.   
Because of the prediction of a diffuse optical extragalactic background 
radiation that is testably different from that in a universe which is not 
time symmetric, we are provided with a two pronged attack on the 
hypothesis of time symmetric 
boundary conditions as a realistic description of our universe: 
the lack of a self-consistent picture of the observed universe, 
and observations of the EGBR.

Second, for completeness it should be mentioned that in the context of
the time neutral generalized quantum mechanics assumed here, there
are further restrictions on the viability of time symmetric models
as a realistic description of our universe which arise from the 
requirements of decoherence and the emergence of approximately classical 
behaviour.\footnote{It is not obvious that boundary conditions of the
character noted above satisfy these restrictions, but if sufficient 
conditions obtain for the RTH to hold then it is at least plausible 
that they do.  This is because such requirements might be expected 
to be more severe the more strongly correlated are the detailed states 
of the expanding and recollapsing eras.  This should become clearer 
in section \ref{sec:tsbc}.} 
These topics are discussed by Gell-Mann and Hartle \cite{TSA}.  

Third, it is important to note that the conclusions of 
section \ref{sec:egbr}\ regarding the EGBR depend essentially 
on the assumed {\it global} homogeneity 
and isotropy, {\it i.e.\ }the ``cosmological principle."   
Thus, while inflation (apparently a generic consequence of 
quantum field theory in a small universe)
can be taken to be a prediction of the assumed boundary conditions if
they are imposed when the universe is sufficiently small, the
popular point of view (somewhat suspect anyway) that inflation provides
an {\it explanation} of homogeneity and isotropy inside the horizon is
not a prediction of the effective theory of the universe considered here;
it is an assumption.  However, the other good things inflation does for us
would still qualify as predictions.   
 
One further point requires mention.  In the conventional picture of
inflation, the matter in the universe is in a pure state (say, some vacuum
state) when small.  After inflation and reheating the matter fields
appear to be, according to local coarse grainings, in thermal equilibrium, 
the states of different fields being highly entangled.  Furthermore, the 
correlations required to infer that the complete state is actually pure 
have been inflated away.  Nevertheless, the quantum state is of course 
still pure \cite{CC}.  As noted previously, pure states are not viable 
candidates for the initial and final conditions.  Therefore, 
in order to fit the conventional picture of inflation into an effective 
theory of the kind considered in this paper, 
the $\rho_{\alpha}$ and $\rho_{\omega}$ defined above must be a local 
description in the sense that they not contain the information required to 
know that the state from which they were inferred was actually pure, 
 {\it i.e.\ }they must coarse-grain this information away.  (The
alternative is to employ an inflationary scenario in which the universe
is not required to be in a pure state.)  The unpleasantness of this 
restriction could be taken as an argument against the use of CPT-related 
boundary conditions in a fundamental quantum cosmological theory.

Finally, for the cognoscenti of generalized quantum 
mechanics \cite[for example]{LesH}, there is a related 
observation that is even more interesting.  Field 
theoretic Hamiltonians are CPT-invariant, so that if the CPT-related 
$\rho_{\alpha}$ and $\rho_{\omega}$ depend only on the Hamiltonian (as for 
instance an exactly thermal density operator does), $\rho_{\alpha}$ and 
$\rho_{\omega}$ are actually {\it equal}.  (The same observation holds for 
merely T-related boundary conditions if the Hamiltonian is T invariant.
However, CPT seems the more relevant symmetry if the Hamiltonian and 
boundary conditions of the considered effective theory of the universe 
arise from some more fundamental theory.)  This is a potential difficulty 
for the sample model 
given above.  A simple extension of a result of Gell-Mann and 
Hartle \cite[section 22.6.2]{TSA} to the case where the background 
spacetime is an expanding universe shows that the only dynamics allowed 
with identical initial 
and final boundary conditions is trivial:  for alternatives allowed in
sets of histories in which probabilities may be assigned at all, the time 
dependence of probabilities for alternative outcomes is essentially 
 {\it independent of the Hamiltonian}, 
and is due only to the expansion of the universe.\footnote{To be more 
precise, consider the projections which appear in sets of histories that 
decohere for some $\rho_{\alpha}=\rho_{\omega}.$  For time independent 
Hamiltonians, time dependence of probabilities for such projections 
arises only if the projections are time-dependent in the Schr\"{o}dinger 
picture.  Due to the expansion, this will be the case for many projections 
onto (otherwise time-independent)
quantities of physical interest in an expanding universe.}
In the present case this may be interpreted as a prediction that 
a universe which is required to be in thermal equilibrium whenever
it is small {\it must remain so when large,} at least in the context 
of the theoretical structure of Gell-Mann and Hartle.  (Classically the 
expectation is the same, so this result can hardly be written off as an 
artifact of the formalism.)  
Now, the evidence suggests that matter in the the early universe 
 {\it was} in local thermal equilibrium.  However, the inhomogeneities 
in the matter required to generate large scale structure must also be 
described by the boundary conditions, and, even in the usual case where 
there is only an initial condition, it is after all gravitational 
condensation which drives the 
appearance of a thermodynamic arrow.\footnote{More precisely, smoothly
distributed matter in thermal equilibrium is {\it not} an equilibrium
state when the gravitational field is included.  Equilibrium states in
the presence of gravitation have clumpy matter in them.}
In the present case (the model with the two-time boundary conditions 
characterized above), the meaning of these observations is that one way 
to break the CPT invariance of the boundary conditions is for the 
specification 
of the deviations from perfect homogeneity and isotropy to be CPT 
non-invariant, in order that the theory admit interesting dynamics.  
As a specific example,
I briefly consider the common circumstance where cosmological matter 
is given a hydrodynamic description.  Scalar-type adiabatic pertubations 
may be completely
specified by, {\it e.g.,} the values of the energy density perturbation
and its (conformal-) time derivative over a surface of constant conformal 
time \cite{MFB}.  (Scalar-type pertubations are the ones of interest, as 
they are the ones which may exhibit instability to collapse.)  
As it is intended here that gravity is treated classically, such a 
specification might come in the form of a probability distribution for 
approximately scale invariant 
metric and energy density fluctuations averaged over macroscopic scales.  
If this probability distribution reflects the underlying FRW homogeneity 
and isotropy (and is thus in particular P-invariant), in order to break 
T-invariance the distribution must distinguish between opposing signs of 
the (conformal-) time derivatives of the energy density perturbations.  
An alternative way to break the equality of CPT-related boundary conditions 
is for the locally observed matter-antimatter asymmetry to extend across 
the 
entire surface of constant universal time.  
(The recollapsing era will then be
antimatter dominated in a CPT symmetric universe, consequently requiring
that there be some mechanism to permit baryon decay.) This possibility may 
appear more attractive, but it is not without significant complications.  
These are addressed briefly at the end of section \ref{sec:amplifications}.  

\subsection{Physics with Time Symmetric Boundary Conditions}
\label{sec:tsbc}

As noted above, the time neutral generalized quantum mechanics of Gell-Mann 
and Hartle \cite{TSA} with CPT-related initial and final density operators
yields a CPT symmetric ensemble of quantum mechanical histories, in the
sense that each history in the collection is accompanied with equal
probability by its CPT-reverse.  The {\it collection} of histories (in
each decohering set) is therefore statistically time symmetric (STS).
Now, a set of histories can be statistically time symmetric without any 
individual history possessing qualitatively similar (if time-reversed) 
characteristics near both ends.  Completely time asymmetric histories can 
constitute a statistically time symmetric set 
if each history in the collection is accompanied by its 
time reverse \cite{TSA,Page2}.  With two-time boundary 
conditions, however, more is required.  That is, there is a 
distinction between a CPT invariant {\it set} of histories, in which
each history in a decohering set is accompanied, with equal
probability, by its CPT reverse, and the considerably more restrictive
notion of a statistically time symmetric universe studied here.  In
virtue of the boundary conditions at both the beginning and the end of
time, probable histories (in a set which decoheres with these boundary
conditions) will in general have both initial and final states which to 
some extent resemble the boundary 
conditions $\rho_{\alpha}, \rho_{\omega}$ \cite{craig95}.

Perhaps the clearest way to understand this is to consider the 
construction 
of a classical statistical ensemble with boundary conditions at
two times.  The probability for each history in the ensemble may be found 
in the following way \cite{Cocke,TSA}.  The boundary conditions are given 
in the form of phase space probability densities for the initial and final 
states of the system.  Pick some initial state.  Evolve it forward to the
final time.  Weight this history by the product of the probabilities that 
it meets the initial and final conditions, and divide by a normalizing 
constant so that all the probabilities sum to one.  Thus, roughly speaking, 
in order for a 
history to be probable in this two-time ensemble both its initial and final 
state must be probable according to the initial and final probability 
densities, respectively.  (For a deterministic classical system, two-time
statistical boundary conditions are equivalent to a boundary condition at 
one time constructed in the obvious way.  In quantum mechanics this is 
no longer true due to the non-commutability of operators at different 
times.  Thus, while
the described algorithm is merely a useful heuristic for understanding the 
implications of two-time boundary conditions {\it classically}, it is more 
essential quantum-mechanically.)  What is not so clear is that the 
resulting probable histories look anything like the probable histories 
in the 
ensemble with just, say, the initial condition.  In general they will not.
Classically, this is merely the statement that the probability measure on a 
space of classical deterministic solutions defined by two-time boundary 
conditions will in general be quite different than the measure defined 
by the initial probability density only.

For classical systems with stochastic dynamics or for quantum statistical
systems, the simple models studied by Cocke \cite{Cocke},
Schulman \cite{Schulman2,Schulman3,Schulman4,Schulman5}, 
and others yield some insight into what is required for evolution near 
either boundary condition to be influenced only by that boundary 
condition.  Here I
merely summarize the intuitively transparent results of this work in the
context of time symmetric boundary conditions.  In particular, the
emphasis will be on boundary conditions which represent ``low entropy"
initial and final states.  

In the absence of a final condition, an initially low entropy state 
generally implies that evolution away from the initial state displays 
a ``thermodynamic arrow" of entropy increase 
(relative to the coarse graining defining the 
relevant notion of entropy; see \cite{Zeh2} for a pertinent overview.)
In an ensemble with two-time, time symmetric, low entropy boundary
conditions, under what conditions will a familiar ``thermodynamic arrow"
appear near either boundary condition?  A pertinent observation is that, 
roughly speaking, in equilibrium all arrows of time disappear.  Therefore, 
if the total time between the imposition of the boundary conditions is much 
longer than the ``relaxation time" of the system to equilibrium (defined as 
the characteristic time a similar system with an initial condition only
takes to relax), it might be expected that entropy will increase in the
familiar fashion away from either boundary condition.  That is,
following the coarse-grained evolution of the system forward from the
initial condition, entropy increases at the expected rate until the
system achieves equilibrium.  The system languishes in equilibrium for a
time, and then entropy begins to decrease again until the entropy
reaches the low value demanded by the final condition.  Thus, this is a 
system in which the ``thermodynamic arrow of time" reverses itself, which
reversal is enforced by the time symmetric, low entropy boundary
conditions.  There is no cause to worry about the coexistence of
opposed thermodynamic arrows.  Histories for which entropy increases away 
from either boundary condition are readily compatible with both boundary 
conditions, and the probable evolutions are those in which, {\it near either 
boundary condition}, there is a familiar thermodynamic arrow of entropy
increase away from the nearest boundary condition.  There, the other
boundary condition is effectively invisible.  Essentially this is
because in equilibrium all states compatible with constraints are
equally probable; the system ``forgets" its boundary conditions.

On the other hand, if there is not time enough for equilibrium to be
reached there must be some reconciliation between the differing arrows of 
increasing entropy \cite{Cocke,PTA,Penrose1,Penrose2,TSA,Schulman5}.
As only histories which satisfy the required boundary conditions are 
allowed, 
it can be anticipated that the statistics of physical processes that would 
ordinarily (in the absence of the final boundary condition) lead to 
equilibration would be affected because histories which were probable with 
an initial condition only are no longer compatible with the final condition.  
The rate of entropy increase is slowed, and in fact the entropy may never 
achieve its maximum value.  In other words, the
approach to equilibrium is suppressed by the necessity of complying with
the low entropy boundary condition at the other end.  More probable in
this two-time ensemble is that the state {\it will continue to resemble the
low entropy boundary conditions.}  The entire evolution is sensitive to the 
presence of {\it both} boundary conditions.

In fact, as noted already, simple models of statistical systems with a 
boundary condition of low entropy at two times bear out these 
intuitively transparent 
expectations \cite{Cocke,Schulman2,Schulman3,Schulman4}.  
Moreover, in the case 
of Markov processes Schulman has demonstrated the described behaviour 
 {\it analytically} \cite{Schulman5}.  This confluence of intuitive 
clarity, and, for simple systems, analytic and (computerized) 
experimental evidence will be taken as suggestive that the significantly 
more complicated physical system considered in this work (matter fields 
in a dynamic universe) behaves in a qualitatively similar fashion when 
burdened with time symmetric boundary conditions.\footnote{Of course, 
the physics must at the completely fine-grained level be consistent 
with whatever dynamics the system obeys, including such 
constraints as conservation laws.  As a moment's reflection on classical 
Hamiltonian systems with two-time phase space boundary conditions reveals, 
this may be a severe constraint!  A purely stochastic dynamics (no
dynamical conservation laws) allows systems great freedom to respect 
the RTH, and conclusions drawn from 
the behaviour of such systems may therefore be misleading.  The
restrictions on sensible boundary conditions in time neutral generalized
quantum mechanics noted in the last paragraphs of section 
\ref{sec:motivations}\ (which arise essentially as a result of the 
requirements of decoherence) are examples of phenomena with no counterpart 
in stochastic systems.  See also the discussion of gravitational collapse 
in section \ref{sec:collapse}.}

The lesson of this section is that the place to look for signs of 
statistical time symmetry in our universe is in physical processes 
with long ``relaxation times," or more generally in any process 
which might couple the expanding and recollapsing eras \cite{TSA}.  
Such processes might be constrained by the presence of a final boundary 
condition.  Obvious candidates include decays of long-lived 
metastable states \cite{Wheeler1,Wheeler2,Schulman3,Schulman5,TSA}, 
gravitational collapse 
\cite[see also \cite{Schulman3}]{Penrose1,Penrose2,Laflamme} 
and radiations \cite{DT,TSA} with great penetrating power such as 
neutrinos, gravitational waves, and possibly 
electromagnetic radiation.  Thus, there are a variety of tests a 
cosmological model with both an initial and a final condition must pass 
in order to provide a plausible description of our universe.

The next two sections discuss the extragalactic background radiation as an 
example of a physical prediction that is expected to be sensitive to the 
presence of a final boundary condition (time symmetrically related to the 
initial.)  I focus on electromagnetic radiation because it is the most 
within our present observational and theoretical grasp, but the essence 
of the discussion is relevant to any similar wave phenomenon.  Finally, 
in section \ref{sec:difficulties}\ I 
offer a few comments on some other issues that need to be addressed 
in any attempt to describe our universe by a model such as the one 
sketched in section \ref{sec:motivations}.

\section{The Opacity of the Future Light Cone}
\label{sec:opacity}
\setcounter{equation}{0}

The aim of this section is to extend arguments of Davies and Twamley 
 \cite{DT} showing that a photon propagating in intergalactic space is 
likely to survive until the epoch of maximum expansion (assuming the 
universe to be closed), no matter how long the total lifetime of the 
universe.  That is, the future light cone is essentially transparent 
over a wide range of frequencies for extinction processes relevant 
in the intergalactic medium (IGM), with an optical depth 
of at most $\tau \sim .01$ at optical frequencies.  The physics in this 
result is that, in cases of physical interest, the dilution of 
scatterers due to the expansion of the universe wins out over the 
extremely 
long path length the photon must traverse.  As a consequence, the 
integrated background of light from galaxies in the expanding phase will 
still be present in the recollapsing phase.  (As explained in section
\ref{sec:egbr}, it 
is this fact which implies that there is an ``excess" EGBR in a time 
symmetric universe that is not associated with galaxies to our past.)  
To show
this, I compute, for a fairly general class of absorption coefficients, the 
optical depth between the present epoch and the moment of maximum expansion.  
For realistic intergalactic extinction processes this optical depth turns 
out to be small.  Indeed, the opacity of the future light cone turns out 
to be dominated by ``collisions" of intergalactic photons with other
galaxies, if they are regarded as completely opaque hard spheres.  Even in
the limit that the lifetime of the universe $T$ becomes infinite 
($\Omega \rightarrow 1$ from above) all of the relevant processes yield
finite optical depths.  As this is essentially the limit of a flat
universe, it is no surprise that extremely simple expressions result.

The results of this section are modest extensions of the work of Davies
and Twamley \cite{DT}.  For the intergalactic extinction mechanisms and
at the frequencies they consider, the formulae for the opacity derived 
here give numbers in agreement
with their results (using the same data for the IGM, of course.)  
The present work is of slightly broader applicability 
in that the opacity is evaluated for 
a fairly general class of frequency dependent extinction coefficients 
(not just for a few specific processes), and its behaviour as the lifetime 
of the 
universe becomes arbitrarily long is determined.  It turns out that quite 
generally, the asymptotic limit is in fact of the order of magnitude of
the {\it upper} limit to the opacity in a closed universe.  These results 
are less general than that of \cite{DT} in that I do not include the 
effects of a cosmological constant (which makes the perturbative analysis 
below significantly more awkward.)  However, a small cosmological constant 
does not effect the qualitative nature of the conclusions of this section.


\subsection{The Future Light Cone Can Be Transparent}
\label{sec:transparent}

Optical depth $\tau$ is defined by
\begin{equation} 
        d\tau = \Sigma \, dl,          
\end{equation}
where $\Sigma$, the linear extinction coefficient, is the fractional
loss of flux per unit (proper) length $l$, and is given microscopically by
\begin{equation} 
                  \Sigma = \sigma n           \label{eq:egbrsigmadef}
\end{equation}
for incoherent scattering from targets with cross section $\sigma$ and
proper number density $n$ (this neglects stimulated emission and scattering
into the line of sight.)  Given $\tau$, the flux density along the line
of sight thus obeys
\begin{equation}  
              i(l) = i_{0}e^{-\tau (l)}.
\end{equation}
Put another way, the probability a photon will propagate a distance $l$
without being absorbed is $ e^{-\tau (l)} $.  For further details, see
for example \cite{RL}.

In order to compute $\tau$ we need $\Sigma (l)$.  In the approximation
(appropriate to the calculation of optical depths between the present and
the moment of maximum expansion) that the universe is exactly described by 
closed, dust-filled Friedmann-Robertson-Walker, it turns out to be helpful 
to trade in the dependence on proper length for time.  The metric is
\begin{equation} 
      ds^{2} = a^{2}\, [ -d\eta^{2} + d\Omega_{3}^{2} ] ,
\end{equation}
where
\begin{equation}  
      d\Omega_{3}^{2} = d\chi^{2} + \sin^{2}\chi \, d\Omega_{2}^{2} 
\end{equation}
is the metric on the unit 3-sphere, and
\begin{equation} 
     dt = a \, d\eta 
\end{equation}
relates the cosmological time (proper time in the cosmological rest
frame) to the conformal time $\eta $.  For dust the time dependence
of the scale factor $a$ can be expressed parametrically as
\begin{equation} 
      a(\eta) = M \, (1-\cos \eta ),     \label{eq:a=}   
\end{equation} 
so that
\begin{equation} 
      t(\eta ) = M \, (\eta - \sin \eta ).   \label{eq:t=} 
\end{equation}    
The lifetime of the universe is then $ T = 2\pi M$.  Here $ M =
\frac{4\pi}{3}\rho a^{3} $ is a constant as the universe expands,
$\rho $ being the mass density of the dust.  $M$ is related to 
more familiar cosmological parameters by 
\begin{equation} 
      M = \frac{1}{H_{0}}\frac{q_{0}}{(2q_{0}-1)^{\frac{3}{2}}}. 
      \label{eq:M}
\end{equation}
Employing the symmetry of the model to take a photon's path as radial,
$ds^{2} = 0$ gives $dl = dt = a \, d\eta $, from which
\begin{equation} 
        d\tau = \Sigma a \, d\eta.    \label{eq:dtau=}
\end{equation} 

The high symmetry of Friedmann-Roberston-Walker also means that nearly
all the relevant physical quantities simply scale as a power of $a$.  
Thus, as will be seen explicitly in the next subsection, it is necessary 
to consider only extinction coefficients of the form
\begin{equation} 
        \Sigma = \Sigma_{0} \, {\left( \frac{a_{0}}{a}\right)}^{p+1}
                                                    , \label{eq:sigma=}  
\end{equation} 
where $a_{0}$ is some fiducial scale factor (conventionally the present
one) and $p$ is a number.

The goal is to compute the optical depth between now and the moment of
maximum expansion.  ``Now" will be taken to be the time $t_{0}$ from the 
big bang to the present.  For absorption coefficients of the form 
(\ref{eq:sigma=}), the optical depth of the future light cone is
\begin{eqnarray}
   \tau   & = & \int_{\tau(t_{0})}^{\tau(T/2)}\! d\tau \nonumber \\
          & = & \int_{\eta_{0}}^{\pi }\! \Sigma \, a \, d\eta \nonumber \\
          & = & \Sigma_{0} \, t_0 \, g_{p}(\eta_{0})     \label{eq:tau=}
\end{eqnarray}
using (\ref{eq:a=}) and (\ref{eq:dtau=}).  Here $\eta_{0}$ is 
the conformal time of the present epoch, and $g_{p}(\eta_{0})$ 
is the dimensionless function
\begin{eqnarray}
  g_{p}(\eta_{0})&  \equiv &  \left( \frac{M}{t_{0}}\right) 
     {\left( \frac{a_{0}}{M}\right)}^{p+1} 
     \int_{\eta_{0}}^{\pi} \frac{d\eta }{(1-\cos \eta )^{p}}  \nonumber \\
              & = &  \frac{(1-\cos \eta_{0} )^{p+1}}{\eta_{0}-\sin\eta_{0}}
                  \int_{\eta_{0}}^{\pi} \frac{d\eta}{(1-\cos \eta )^{p}}  
                                         .             \label{eq:g=}
\end{eqnarray}

For integral and half-integral $p$ explicit evaluation of $g_{p}(\eta_{0})$ 
is possible, but not terribly illuminating.  In the limit that the total
lifetime of the universe $T$ is very long compared to $t_{0}$, however,
simple expressions for any $p$ result.  (This is no surprise as the
results must approach those of a flat universe.)  One straightforward
procedure involves inverting (\ref{eq:t=}) to get a power series in 
${\left(\frac{t_{0}}{M}\right)}^{\frac{1}{3}}$ for $\eta_{0}$,  and using 
this 
to evaluate the asymptotic behaviour of $g_{p}(\eta_{0})$ as $M$ becomes
large relative to $t_{0}$, which is held fixed.  ($T=2\pi\, M \gg t_{0}$
corresponds to $\eta_{0} \ll 1$.)  It is then tedious but straightforward
to show that 
\begin{equation}
g_{p} \sim 
\left\{   \begin{array}[c]{ll}
\frac{3}{2p-1} \left[ 1+\frac{p+1}{10(2p-3)}
 {\left( \frac{6t_{0}}{M} \right) }^{\frac{2}{3}} \right]     
            & \mbox{$ p > \frac{1}{2} \ ; \ p \not= \frac{3}{2} $} \\
\frac{3}{2}
\left[ 1-\frac{1}{12}{\left( \frac{6t_{0}}{M} \right)}^{\frac{2}{3}} 
\ln\left(\frac{6t_{0}}{M}\right) \right] 
                   & \mbox{$p = \frac{3}{2}$}\\
\ln\left(\frac{M}{t_{0}}\right)
\left[ 1-\frac{3}{40}{\left(\frac{6t_{0}}{M}\right) }^{\frac{2}{3}} 
   \right]
                   & \mbox{$p=\frac{1}{2} $}\\
3\pi \frac{\Gamma (1-2p)}{\Gamma^{2} (1-p)}
 {\left( \frac{M}{6t_{0}}\right) }^{\frac{1-2p}{3}} - \frac{3}{1-2p}
                   & \mbox{$-\frac{1}{2} < p < \frac{1}{2}$} \\ 
12{\left( \frac{M}{6t_{0}}\right) }^{\frac{2}{3}} - \frac{9}{5}
                   & \mbox{$ p = -\frac{1}{2}$} \\ 
3\pi \frac{\Gamma (1-2p)}{\Gamma^{2} (1-p)}
 {\left(\frac{M}{6t_{0}}\right) }^{\frac{1-2p}{3}}
\left[ 1-\frac{p+1}{20}{\left( \frac{6t_{0}}{M}\right) }^{\frac{2}{3}} 
           \right]
                   & \mbox{$ p < -\frac{1}{2}$} 
                                           \end{array} \right.
                                               \label{eq:g}
\end{equation}
In each case only the leading order correction in $\frac{t_{0}}{M}$ 
has been retained.   The most important thing to notice is that 
for $p > \frac{1}{2}$,
$g_{p}$ is perfectly finite even as the lifetime of the universe becomes
arbitrarily big, and as $\frac{t_{0}}{M}$ becomes very small
the opacity converges to the value it would have in a flat universe.
It is clear that for $p > \frac{3}{2}$ the $\Omega_{0} = 1$ result is a
local lower limit on the opacity of the future light cone, as may be
verified directly also from the available exact results.  However, for
reasonable $p$ the corrections to the flat universe result are only a 
factor of order one.  (In fact, examination of the exact results reveals 
that the {\it maximum} of $g_{p}$ as one varies $\frac{t_{0}}{M}$ is at 
most $20\%$ larger than the flat universe result for $p \sim {\rm few}$.
This is a good thing, because unless $\Omega$ is fairly close to one, 
$\frac{t_{0}}{M}$ is not a particularly small parameter!  
In terms of familiar cosmological parameters,
\begin{equation}
         \frac{t_{0}}{M}= \left[ \cos^{-1}({q_{0}}^{-1}-1)
                          -{q_{0}}^{-1}(2q_{0}-1)^{\frac{1}{2}}\right].)
                                         \label{eq:t/M}
\end{equation}
Physically, what's at work is the competition between the slower
expansion rates of universes with larger $\Omega_{0}$'s, which tends to
increase the opacity because the scattering medium isn't diluted as
rapidly,  and the decrease in the opacity due to the shortened time
between the present epoch and the moment of maximum expansion.\footnote{It
will be noticed by combining (\ref{eq:M}) and (\ref{eq:t/M}) that taking
the limit $\Omega_{0} \rightarrow  1$ holding $t_{0}$ fixed requires $H_{0}$ 
to vary as well, converging to the flat universe relation $H_{0} =
\frac{2}{3t_{0}}$.  It is possible to repeat the entire analysis holding the
observable quantity $H_{0}$ fixed instead of $t_{0}$ (for this purpose
the more standard redshift representation is more useful than that in
terms of conformal time used above), but unsurprisingly the conclusions 
are the same:  the opacity is always finite for $p > \frac{1}{2}$;
as $\Omega_{0}$ approaches one, the opacity approaches the flat universe 
result; and the {\it maximum} opacity for reasonable $p$ is only a factor 
$ \ltwid 1.2$ 
times the flat universe result.  The resultant opacities are of course 
related in these limits via $t_{0} = \frac{2}{3H_{0}}$.  Similarly, 
it is possible to perform a related analysis of more complicated 
extinction coefficients than 
(\ref{eq:sigma=}), for example incorporating the exponential behaviour 
encountered in free-free absorption (see (\ref{eq:Sff})) or in modeling 
evolving populations of scatterers with, for instance, a Schecter function 
type profile.  However, these embellishments are not required in the sequel, 
and the techniques are tedious and fairly ordinary,  
so space will not be taken to describe them here.}

To summarize, all of the processes relevant to extinction 
in the intergalactic medium have extinction coefficients that 
can be bounded above by a coefficient
of the form (\ref{eq:sigma=}) with $p > \frac{1}{2}$.  
Using the limiting relationship 
$t_{0} = \frac{2}{3H_{0}}$, we have from (\ref{eq:tau=}) and (\ref{eq:g}) 
the simple result that for these processes, the upper limit to the opacity 
between the present epoch and the moment of maximum expansion, no matter 
how long the total lifetime of the universe, is of order
\begin{equation}
              \tau = \frac{2}{2p-1}\frac{\Sigma_{0}\, c}{H_{0}}.
                                               \label{eq:tau}
\end{equation}
(I have returned to conventional units in this formula.)

\subsection{The Opacity of the Future Light Cone}
\label{sec:flc}

In this section I apply the asymptotic formula (\ref{eq:tau}) for the upper 
limit to the optical depth of the FLC in a 
long-lived universe to show that if our universe is closed, photons escaping 
from the galaxy are (depending on their frequency) likely to survive into
the recollapsing era.   That is, the finite optical depths computed in
the previous section are actually small for processes of interest in the
intergalactic medium (IGM).  For simplicity, I focus on photons
softer than the ultraviolet at the present epoch; the cosmological 
redshift makes it necessary to consider absorption down to very low 
frequencies.

It is important to note that in employing standard techniques for
computing opacities the effects of the assumed statistical time symmetry
of the universe are being neglected.  As discussed in section
\ref{sec:tsbc}\ and in section \ref{sec:difficulties}, when 
the universe is very large the thermodynamic and gravitational behaviour of
matter will begin to deviate from that expected  were the universe not
time symmetric.  Due to the manifold uncertainties involved here it is
difficult to approach the effects of time symmetric boundary conditions
on the opacity of the
future light cone with clarity.\footnote{For example, how is scattering of 
light by a ``thermodynamically reversed" medium to be treated, as when
light from the expanding era reaches the intergalactic medium in the
recollapsing phase?  The standard account assumes incoherent scattering.
Thus a laser beam shone on a plasma is diffused.  Time-reversing this
description yields {\it extremely} coherent scattering from the plasma
which reduces its entropy.  Thus scattering or absorption of light 
correlated
with sources (such as galaxies) in the expanding phase by material in the
recollapsing phase appears to require entropy reducing (according to the
observers of the recollapsing era) correlations in the matter there, in
contradiction with the presumed local thermodynamic arrow (and  
with the RTH), in order to yield what there appears as emission.
This is just the sort of detailed connection between the expanding and
recollapsing eras which would lead one to expect physical predictions in
a time symmetric model, even very near one of the boundary conditions,
to be very different than those in a model with an initial condition
only (section \ref{sec:tsbc}).  This complication is closely related 
to the difficulty, mentioned in section \ref{sec:retardation}, in 
deriving the retardation of radiation in a universe
which is time symmetric and in which the future light cone is transparent.}
I shall assume they are not such as to increase it.
This is reasonable as the dominant contribution to the opacity comes
when the universe is smallest, where in spite of the noted complications 
the RTH is assumed to hold.

What are the processes relevant to extinction of photons in the
intergalactic medium?  Because the IGM appears to consist in hot,
diffuse electrons, and perhaps a little dust \cite{BFR},  extinction
processes to include are Thomson scattering, inverse bremsstrahlung
(free-free absorption), and absorption by dust.  In addition, absorption
by material in galaxies (treated as completely black in order to gauge 
an upper limit) is important.  These processes will treated in turn.  
(A useful general
reference on all these matters is \cite{RL}.)  The conclusion will be
that while absorption by galaxies and Thomson scattering are most 
significant above the radio, none of these processes pose a serious 
threat to a photon that escapes from our  galaxy.   This confirms the 
results of Davies and Twamley \cite{DT}, who however did not consider the
possibly significant interactions with galaxies.
Consequently I will be brief.  Some results of Davies and Twamley
regarding absorption mechanisms which may be important when the
universe is very large and baryons have had time to decay are
quoted at the end of this section.   These do not appear to be
significant either.

(For high energy photons Compton scattering, pair production,
photoelectric absorption by the apparently very small amounts of neutral
intergalactic hydrogen, and interactions with CMBR photons will be
important, but as none are significant below the ultraviolet I do not 
discuss them here.  All can be treated by the same methods as the 
lower energy processes.)

To begin, following Davies and Twamley \cite{DT}, I quote 
Barcons {\it et al.\ }\cite{BFR} on current beliefs regarding the state 
of the IGM in the form
\begin{eqnarray}
n_{{\rm H_{II}}} 
           & = &  \delta \, n_{0} \, {\left( \frac{a_{0}}{a} \right) }^{3}
                                              \nonumber \\
T_{{\rm H_{II}}}
           & = & \epsilon \, T_{0} \, {\left( \frac{a_{0}}{a} \right) }^{2}
                                              \label{eq:IGM}
\end{eqnarray}
where
\begin{eqnarray}
n_{0} & = & 1.12 \, h^{2} \, 10^{-7} \, {\rm\,cm}^{-3}, \ \  
                                           \delta \in (1,10) \nonumber \\
T_{0} & = & 10^{4} \, {\rm\,K} ,                        \ \ 
                                             \epsilon \in (1,10^{3})
                                                      \label{eq:IGMparams}
\end{eqnarray}
with the values $\delta = \epsilon = 1$ somewhat preferred by the
authors.  In addition, the present upper limit on a smoothly distributed
component of neutral hydrogen is about 
$n_{{\rm H_{I}}} < 10^{-12} \, {\rm\,cm}^{-3}$.  
Thus, the intergalactic medium consists in hot (but
non-relativistic) electrons, protons, and essentially no neutral
hydrogen.  The lack of distortions in the microwave background indicates
its relative uniformity, at least to our past.  From now on, $n$ and $T$ 
simply will be used to refer to the number density and temperature of 
intergalactic electrons.

Finally, very little is known about a possible diffuse component of
intergalactic dust \cite{BFR,WR}, except that there is probably very
little of it.  Most dust seems to be clumped around galaxies.  Therefore
I will ignore possible extinction due to it, subsuming it into the ``black
galaxy" opacity.  Davies and Twamley \cite{DT} make some
estimates for one model for the dust, finding its contribution to the
opacity insignificant.  At any rate, models for the absorption
coefficient due to dust \cite{Peebles,HW} all give a cross section $\sigma$ 
that falls with increasing wavelength, $\sigma \sim 1/{{\lambda}^{q}}$ with 
$1 \leq q \leq 4$, so that 
$\Sigma = \sigma \, n \propto {\left( \frac{a_{0}}{a}\right)}^{q+3}$
(neglecting of course a clumping factor expressing the fact that
clumping decreases the opacity.)  Thus 
$p_{{\rm dust}} = q + 2 > \frac{1}{2}$, 
the dust opacity is bound to be finite, and with a small present 
density of diffuse dust it is not surprising to find its contribution 
to be small.

Before considering the optical depth due to interactions with
intergalactic electrons, I will show that it is reasonable to
approximate that most photons escaping our galaxy will travel freely
through intergalactic space.  That is, few photons will end up running
into another galaxy.

\subsubsection*{Collisions with Galaxies} 
\label{sec:collisions}

Drastically overestimating the opacity due to galaxies by pretending
that any photon which enters a galaxy or its halo will be absorbed by
it (the ``black galaxy" approximation), and taking the number of 
galaxies to be constant,
\begin{eqnarray}
\label{eq:hardspheresigma}
\Sigma_{{\rm gal}} & = & \sigma \, n  \nonumber \\
         & = & \sigma \, n_{0} \, {\left( \frac{a_{0}}{a} \right) }^{3},
\end{eqnarray}
where $\sigma$ is the cross-sectional area of a typical galaxy and 
$n_{0}$ is their present number density.  Thus from (\ref{eq:tau}), the
upper limit on the opacity due to collisions with galaxies is
\begin{equation}
       \tau = \frac{2}{3}\, \frac{\sigma \, n_{0} \, c}{H_{0}}.
\end{equation}
As noted above, this is finite (even as the lifetime of the universe
becomes very large) because the dilution of targets due to
the expansion of the universe is more important than the length of the
path the photon must traverse.

Notice that assuming target galaxies to be perfectly homogeneously
distributed only overestimates their ``black galaxy" opacity.  Volume
increases faster than cross-sectional area, so clustering reduces the
target area for a given density of material.  As galaxy clustering is
not insignificant today and will only increase up to the epoch of
maximum  expansion even in a time symmetric universe, the degree of
overestimation is likely to be significant.

Taking 
$n_{0}\sim .02\, h^{3}\, {\rm\,Mpc}^{-3}$, 
$\sigma = \pi\, r_{{\rm gal}}^{2}$ 
(where $r_{{\rm gal}} \sim 10^{4}\, h^{-1}\, {\rm\,pc}$), and 
$H_{0} \sim \frac{1}{3} \cdot 10^{-17}\, h\, {\rm\,s}^{-1}$ (here
$.4 < h < 1 $ captures as usual the uncertainty in the Hubble constant)
gives the upper limit
\begin{equation}
                \tau \sim .01.                 \label{eq:HS}
\end{equation} 
This can be interpreted as saying that at most about one percent of the
lines of sight from our galaxy terminate on another galaxy 
before reaching the recollapsing era.  By time symmetry, neither do 
most lines of sight connecting the present epoch to its time-reverse.

\subsubsection*{Thomson Scattering}
\label{sec:thomson}

Use of the Thomson scattering cross section 
$ \sigma_{{\rm T}} = \frac{8\pi}{3}r_{0}^{2} = 6.65 \cdot 10^{-25}\, 
                                                {\rm\,cm}^{-2}$
is acceptable for scattering from non-relativistic electrons for any 
photon softer than a hard X-ray ($\hbar \omega \ll mc^{2}$).  
Thus, for the frequencies I will consider, 
$\Sigma_{{\rm T}} = \sigma_{{\rm T}} n$ 
will suffice, giving
\begin{eqnarray}
   \tau_{{\rm T}} & = & 
             \frac{2}{3}\, \frac{\delta \sigma_{{\rm T}} n_{0} c}{H_{0}}
                                                         \nonumber \\ 
                  & = & 4.7 (\delta h) 10^{-4}.              \label{eq:T}
\end{eqnarray}
Recalling that $\delta$ is at worst one order of magnitude, it is clear
that Thomson scattering is not signficant for intergalactic
photons \cite{DT}.  It is perhaps worth mentioning that quantum and
relativistic effects only tend to decrease the cross section at higher
energies.  More significant for the purposes of this investigation is the
observation that, at the considered range of frequencies, 
Thomson scattering 
does not change a photon's frequency, merely its direction.  Thus
Thomson scattering of a homogeneous and isotropic bath of radiation by a
homogeneous and isotropic soup of electrons has {\it no effect} as
regards the predictions of section \ref{sec:egbr}.\footnote{Were it 
significant, it would however be a means of hiding the 
 {\it information} contained in the background.}

\subsubsection*{Inverse Bremsstrahlung}
\label{sec:bremsstrahlung}

Even less significant than Thomson scattering for frequencies of
interest is free-free absorption by the IGM \cite{DT}.  From, 
 {\it e.g.\ }\cite{RL}, the linear absorption coefficient for scattering
from a thermal bath of ionized hydrogen is 
\begin{eqnarray}
\Sigma_{{\rm ff}} & = & \frac{2e^{6}}{3m\hbar c} {\left( \frac{2}{3\pi km}
\right) }^{\frac{1}{2}} n^{2}T^{-\frac{1}{2}} \nu^{-3} \overline{g}(b) 
   (1-e^{-b}) \nonumber \\
&=&3.7\cdot 10^{8} n^{2} T^{-\frac{1}{2}} \nu^{-3} 
         \overline{g}(b)(1-e^{-b}) {\rm\,cm}^{-1}. 
                                \label{eq:Sff}
\end{eqnarray}
in cgs units.
Here $b \equiv \frac{h\nu}{kT}$, the factor $e^{-b}$ contains the effect
of stimulated emission, and $\overline{g}(b)$ is a ``Gaunt factor"
expressing quantum deviations from classical results.  It is a
monotonically decreasing function of $b$ which is of order one in the
optical ({\it cf.\ }\cite{RL} for a general discussion and 
some references.)  As
\begin{eqnarray}
    b &=& \frac{h\nu}{kT}                    \nonumber \\
      &=& \frac{h\nu_{0}}{kT_{0}} \left( \frac{a}{a_{0}}\right)  \nonumber
\end{eqnarray}
increases as the universe expands, taking $\overline{g}(b) = g_{0}$, a
constant of order one, will only overestimate the opacity.  Similarly,
following \cite{DT} in dropping the stimulated emission term will yield
an upper limit to the free-free opacity.  With $\epsilon = 1$,
$\frac{h\nu_{0}}{kT_{0}} = 1$ when $\nu_{0} \sim 10^{14}\, {\rm\,s}^{-1}$, 
so stimulated emission will only lead to a noticeable reduction in 
$\Sigma_{{\rm ff}}$
well below the optical.  (Actually, methods similar to that employed in
section \ref{sec:transparent}\ can be employed to calculate this term, 
but as $\tau_{{\rm ff}}$ will turn out to be insignificant even 
neglecting it there is no need to go into that here.)

With these approximations,
$$
\Sigma_{{\rm ff}}  \approx  
               (4.6 \cdot 10^{-8}) g_{0}\delta^{2}\epsilon^{-\frac{1}{2}} 
               h^{4} \nu_{0}^{-3} {\left( \frac{a_{0}}{a} \right) }^{2}, 
$$
and thus
\begin{eqnarray}
  \tau_{{\rm ff}} & = & 2 \frac{\Sigma_{0} c}{H_{0}}    \nonumber \\
&=&8.6\cdot 10^{20}h^{3} g_{0}\delta^{2}\epsilon^{-\frac{1}{2}} 
\nu_{0}^{-3}.  \label{eq:ff}
\end{eqnarray}
Recalling that $\delta = \epsilon = 1 $ seem likely physical values, and
noting that $\delta^{2}\epsilon^{-\frac{1}{2}} \ltwid 10^{2}$ at worst, 
taking $h^{3} g_{0}\delta^{2}\epsilon^{-\frac{1}{2}} = 1$ is not 
unreasonable for an
order of magnitude estimate.  Thus $\nu_{0} \sim 10^{7} {\rm s}^{-1}$ 
(long radio) is required to get $\tau_{{\rm ff}} \sim 1$.  Since 
$\tau_{{\rm ff}} \propto \nu_{0}^{-3}$ it drops sharply for photons 
with present frequency above that.  For instance, at 5000\AA
$$ \tau_{{\rm ff}} =  g_{0}\delta^{2}\epsilon^{-\frac{1}{2}} 10^{-24}, $$
and inverse bremsstrahlung is completely negligible.

\subsubsection*{The Far Future}
\label{sec:future}

Finally, I mention that Davies and Twamley \cite{DT} consider 
what happens if 
baryons decay in a long lived universe.  Following the considerations 
of \cite{PM}, they conclude that the positronium ``atoms" which will 
form far in the future (when the universe is large) remain transparent 
to photons with
present frequencies in the optical. This is because the redshifted photons
haven't enough energy to cause transitions between adjacent Ps energy
levels.  Similarly, if in the nearer future the electrons and protons in
the IGM recombine to form more neutral hydrogen, this will also be
transparent at the considered frequencies. 


\section[The EGBR in Time Symmetric Universe]{Extragalactic 
Background Radiation in a \protect\\ 
Statistically Time Symmetric Universe}
\label{sec:egbr}
\setcounter{equation}{0}

\subsection{Lower Limit to the Excess Optical EGBR}
\label{sec:limit}

The goal of this section is to explain why, in a 
statistically time symmetric
universe (such as one with the CPT-related boundary conditions discussed 
in section \ref{sec:motivations}), the optical extragalactic background 
radiation should be at least twice
that expected in a universe which is not time symmetric, and possibly 
considerably more.  Thus, assuming consistency with the 
RTH ({\it i.e.\ }the
predictive assumption that physics near either boundary condition is 
practically insensitive to the presence of the other boundary condition, 
 {\it cf.\ }section \ref{sec:tsbc}), it is possible to discover 
 {\it experimentally}
whether our universe is time symmetric.  Section \ref{sec:observations}\
compares this prediction with present observations, concluding that the 
minimal
prediction is consistent with upper limits on the observed optical EGBR.
However, better observations and modeling may soon challenge even this
minimal prediction.

At optical wavelengths, the isotropic bath of radiation from sources outside 
our galaxy is believed to be due almost exclusively to galaxies on our past 
light cone \cite[are some good general references]{Peebles,Deep,IAU,EGBR}.
There is no other physically plausible source for this radiation.  
In a model universe with time symmetric boundary conditions, however, there 
must in addition be a significant quantity of radiation correlated with the 
time-reversed galaxies which will exist in the recollapsing era, far to our 
future \cite{DT,TSA}.  The reason for this is that light from our galaxies 
can propagate largely unabsorbed into the recollapsing phase no matter how 
close to open the universe is, as shown in \cite{DT} and in section
\ref{sec:opacity}.
This light will eventually arrive on galaxies in the recollapsing phase,
or, depending on its frequency, be absorbed in the time-reversed equivalent
of one of the many high column density clouds (Lyman-limit clouds and damped 
Lyman-$\alpha$ systems) present in our 
early universe \cite{Peebles,Deep,QSO},
in the intergalactic medium, or failing that, at the time-reversed
equivalent of the surface of last scattering.  This will appear to observers 
in the recontracting phase as emission by one of those sources sometime in 
their galaxy forming era.  Since future galaxies, up to high time-reversed 
redshift, 
occupy only a small part of the sky seen by today's (on average) 
isotropically 
emitting galaxies, much of the light from the galaxies of the expanding
phase will proceed past the recontracting era's galaxies.  Thus most of 
this
light will be absorbed in one of the other listed media.  
Because of the assumption of global homogeneity and isotropy, the light 
from the entire history of galaxies in the
expanding phase will constitute an isotropic bath of radiation to
observers at the time-reverse of the present epoch that 
is {\it in addition} to the light from the galaxies to {\it their} past.  
By time symmetry, there will be a similar contribution to our EGBR 
correlated with galaxies which will
live in the recollapsing phase, over and above that due to galaxies on
our past light cone.  To us this radiation will appear to arise in
isotropically distributed sources {\it other} than galaxies.  This 
picture of a transparent, time symmetric universe is illustrated in 
figure 1.  
\begin{figure}[ht]
\label{fig:egbr}
\begin{center}
\epsfig{file=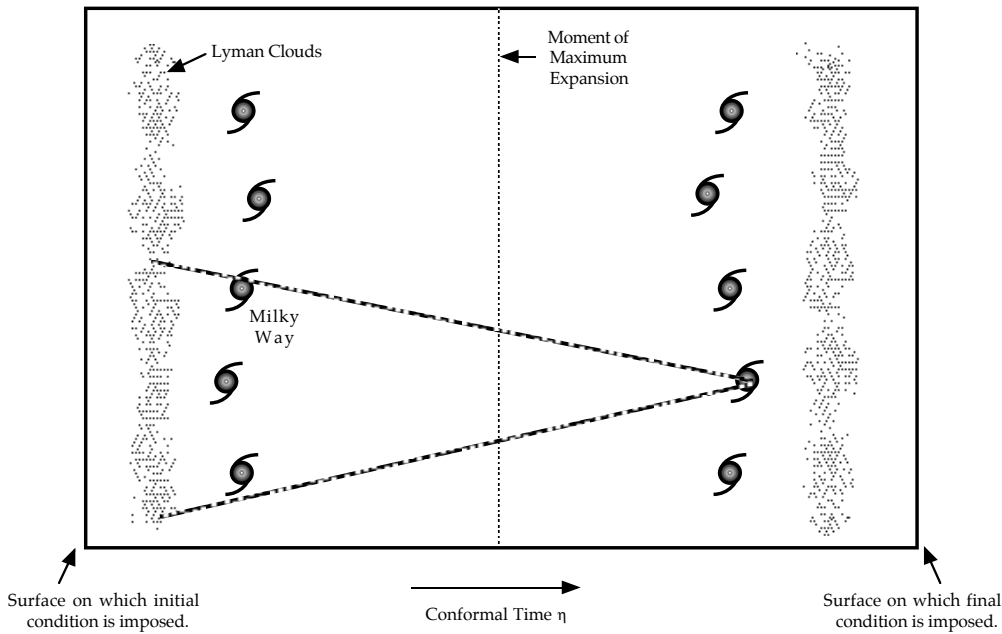}
\end{center}
\caption{Schematic representation of the
origin of the ``excess'' extragalactic background light correlated
with the thermodynamically reversed galaxies of the recollapsing era.
The model is a Friedmann-Robertson-Walker universe equipped with time
symmetric boundary conditions requiring the universe to be smooth
and in local thermodynamic equilibrium whenever it is small.}
\end{figure}

A lower limit to this excess background can be obtained by considering how 
much light galaxies to our past have emitted already ({\it cf.\ }section
\ref{sec:observations}).  
According to observers at the time reverse of the present epoch,
this background will (in the absence of interactions) retain its 
frequency spectrum and energy density because the size of the universe is 
the same.  Thus, by time symmetry, at a {\it minimum} the predicted optical
EGBR in a universe with time symmetric boundary conditions is twice that 
expected in a universe in which the thermodynamic arrow does not reverse.
If much of the luminous matter in galaxies today will eventually be burned 
into radiation by processing in stars or galactic black holes, the total 
background radiation correlated with galaxies in the expanding phase could 
be several orders of magnitude larger, a precise prediction 
requiring a detailed understanding of the future course of galactic 
evolution \cite{DT}.  Several other processes may also contribute 
significant excess backgrounds.  These topics are discussed further below.

\subsubsection*{The Prediction}

A number of points in the summary argument above require amplification.
First, however, I summarize the {\it minimal}\footnote{By ``minimal" I mean 
the lower limit in each band provided by taking the integrated background of 
light from galaxies in the expanding era to be only that which has 
 {\it already}
been emitted up to the present epoch.  In the absence of absorption, by time 
symmetry this is the minimum background at the present epoch that must be 
correlated with galaxies in the recontracting phase, as in the previous 
paragraph.} predictions for the ``excess" extragalactic background 
({\it i.e.\ }radiation from non-galactic sources to our past 
that is correlated with time-reversed galaxies) in bands 
for which the future light cone is transparent:
\begin{itemize}
\item isotropy:  the ``future starlight" should appear in the comoving frame
as an approximately isotropic background.  This conclusion depends 
crucially on the assumed global validity of the cosmological principle.
\item energy density: comparable to the present energy density in starlight 
due to the galaxies on our past light cone.  This assumes the future light
cone (FLC) is totally transparent. 
\item spectrum: similar to the present spectrum of the background starlight 
due to galaxies on our past light cone.  Again, neglect of further emissions
in the expanding phase makes this, by time symmetry, a lower limit in
each band.  This conclusion relies on the assumption of a transparent FLC 
in part to the extent that this implies a paucity of standard astrophysical
mechanisms for distorting spectra.
\end{itemize}

Thus, at for instance optical frequencies, time symmetry requires an
isotropic extragalactic background at least twice that due to galaxies on
our past light cone alone.\footnote{To be totally accurate, the quantity
of radiation absorbed or scattered into another band between the present 
epoch and its time reverse should be subtracted.  However, the upper limit 
to the total FLC opacity (due to anticipated processes) computed in section
\ref{sec:opacity}\ was of order $10^{-2}$, mostly due to a liberal 
(``black galaxy") assessment of the rate of 
interception of photons by galaxies, and I will therefore neglect 
such losses.  Further, it is worth remembering that processes like 
Thomson scattering do not destroy photons or change their frequency, 
but only scatter them.  Thus mere 
scattering processes may introduce isotropically distributed (via the 
cosmological principle) fluctuations in the background, but not change its 
total energy.  Similarly, line or dust absorption usually result in 
re-radiation of the absorbed photons, conserving the total energy in an 
isotropic background (if the size of the universe doesn't change much before 
the photons are re-radiated), if not the number of photons with a given 
energy.}
The potentially far greater background predicted (by time symmetry) if 
further emissions in the expanding phase are accounted for is a subject 
taken up in the sequel.

\subsubsection*{Consistency with the RTH?}

Before proceeding, a comment on the consistency of this picture is 
in order.  
As the ``excess" radiation is correlated with the detailed histories of
future galaxies, the transparency
of the future light cone does not appear consistent with the predictive 
assumption (the RTH) that physics in the expanding era should be 
essentially
independent of the specifics of what happens in the recontracting phase.
At a minimum, if the model is to be at all believable it is legitimate to 
demand that the required radiation appears to us to arise in sources in a 
fashion consistent 
with known, or at least plausible, astrophysics.  Thus it may be 
that given a transparent FLC, the only viable picture of a time symmetric 
universe is one in which the radiation correlated with future galaxies {\it 
``should" be there anyway, i.e.\ }be predicted also in some reasonable model 
of our universe which is {\it not} time symmetric, and consequently not
be ``excess" radiation at all, but merely optical radiation arising in
non-galactic sources during (or before) the galaxy forming era.  
On the basis of present knowledge this does not describe our universe.  
The presence of the radiation required by time symmetry and the 
transparency of the FLC appears to be in significant disagreement 
with what is known about our galaxy 
forming era, as will become apparent below.  

Were it the case for our universe that non-galactic sources provided a
significant component of the optical EGBR, the difficulties with 
time symmetric
boundary conditions would from a {\it practical} point of view be less 
severe.  It is true that the non-galactic sources emitting the 
additional isotropic background would have to do so in just such a way 
that the radiation contain the correct spatial and spectral correlations 
to converge on 
future galaxies at the appropriate rate.  This implies a distressingly 
detailed connection between the expanding and recontracting phases. 
However, if the emission rate and spectrum were close to   
that expected on the basis of conventional considerations these 
correlations (enforced by the time symmetric boundary conditions) 
would likely be wholly 
unobservable in practice, existing over regions that are not causally 
connected until radiation from them converges onto a future 
galaxy \cite{DT},
and thus not visible to local coarse grainings (observers) in the expanding 
era.  In any event, the meaning of the transparency of the FLC is that
starlight is by no means a ``short relaxation time process."
Of course, on the basis of the models discussed in section \ref{sec:tsbc}, 
perhaps the conclusion to draw from this apparent inconsistency between the 
RTH and the transparency of the FLC
should rather be, that physical histories would unfold in a fashion 
quite different from that in a universe in which $\rho_{\omega}=1$, namely, 
in such a way that such detailed correlations would never be required in 
the first place.  The very formation of stars might be suppressed 
({\it cf.\ }section \ref{sec:difficulties}).  
Be that as it may, to the extent that the universe to our
past is well understood, there are {\it no} sources that could plausibly 
be responsible for an isotropic optical background comparable to that 
produced by galaxies.  Such an additional background, required in a 
transparent, time symmetric universe, requires significant, observable 
deviations from established astrophysics.  This is in direct contradiction 
with the RTH.  Thus the entire structure in which an
additional background is predicted appears to be both 
internally inconsistent (in that it is inconsistent with the postulate 
which allows predictions to be made in the first place), and, 
completely apart from observations of the EGBR 
discussed in section \ref{sec:observations}, 
inconsistent with what is known about our galaxy forming
era.  This should be taken as a strong argument that our universe does not
possess time symmetric boundary conditions.  Nevertheless, in order to 
arrive at this conclusion it is necessary to pursue the consequences of 
assuming the consistency of the model.  The situation may be stated thus:
 {\it either} our universe is not time symmetric, {\it or} there is an 
unexpected contribution to the optical EGBR due to non-galactic sources to 
our past (and there are indeed detailed correlations between the expanding 
and recontracting 
eras), {\it or} perhaps our future light cone is not transparent after all.  
This latter possibility, perhaps related to the considerable uncertainty 
regarding the state of the universe when it is large, 
seems the last resort for a consistent time symmetric model of our universe.

\subsection{Amplifications}
\label{sec:amplifications}

It is now appropriate to justify further some of the points made in
arriving at the prediction of an excess contribution to the EGBR.
Claims requiring elaboration include: i) most of the light from the
expanding era's galaxies won't be absorbed by the galaxies of the
recontracting phase, and {\it vice-versa}; ii) it will therefore be
absorbed by something else, and this is inconsistent with the early
universe as presently understood; iii) a detailed understanding of the
old age and death of galaxies, as well as other processes when the
universe is large, may lead to a predicted EGBR in a time symmetric
universe that is orders of magnitude larger than the minimal prediction
outlined above.  I deal with these questions in turn.

\subsubsection*{The ``Excess" EGBR is Not Associated with 
Galaxies to Our Past}

A photon escaping our galaxy is unlikely to encounter another galaxy
before it reaches the time reverse of the present epoch.  In fact, as
shown in section \ref{sec:opacity}, galaxies between the present epoch
and its time reversed 
equivalent subtend, at most, roughly a mere $ 2 \times .01 = 2\% $ 
of the sky (\ref{eq:HS}) (neglecting curvature and clumping.)
In light of the present lack of detailed information about our galaxy 
forming era (and via time symmetry its time-reverse), a photon's fate 
after that is more difficult to determine.  A straightforward extrapolation 
of the results of 
section \ref{sec:opacity}\ (or {\it cf. e.g.\ }section 13 of Peebles 
\cite{Peebles}) shows that the optical
depth for encounters with galaxies of the same size and numbers as today is
$\tau \sim .01 (1+z)^{\frac{3}{2}}$ between a redshift of $z$ and today.  
Again this neglects curvature (hence overestimating $\tau$) and clumping 
(which now underestimates $\tau$.)  Assuming the bright parts of galaxies 
form at $z \sim 5$, this gives only $\tau \sim .14$, and the sky isn't 
covered with them until $z \sim 20$.  This however is roughly at the upper
limit on how old galaxies are thought to be. 
On the other hand, examination of quasar spectra 
(out to $z \sim 5$) show that most lines of sight pass through many clouds
of high column densities of hydrogen called Lyman-$\alpha$ forest clouds 
and, 
at higher densities, damped Lyman-$\alpha$ systems.   
(Peebles \cite{Peebles} 
is a useful entry point on all of these matters, as is \cite{Deep}.
\cite{QSO} are the proceedings of a recent conference concerned with 
these Lyman systems.)  The highest density clouds 
may be young galaxies, but if so galaxies were more diffuse in the past 
as the observed rate of interception of an arbitrary line of sight with 
these clouds is a factor of a few or more greater than that based  
on the assumption that 
galaxy sizes are constant.  (Obviously, this would not be too surprising.) 
For instance, for the densest clouds Peebles \cite[section 23]{Peebles} 
relates the approximate formula
$$     \frac{dN}{dz} = 0.3 \, {\Sigma_{20}}^{-0.46}    $$
for the observed interception rate per unit redshift of a line of
sight with a cloud of column density greater than or equal to 
$\Sigma_{20}$ (in units of $10^{20} {\rm\,cm}^{-2}$),
in a range of redshifts about $z=3$.   For Lyman-$\alpha$ forest clouds, 
$\Sigma \gtwid 10^{14} {\rm\,cm}^{-2}$, 
the interception rate is considerably higher.
(For some models see \cite{Tytler,Sargent}.)
Thus an arbitrary line of sight arriving on our galaxy from a redshift 
of five, say, is likely to have passed through at least one cloud of column 
density comparable to a galaxy, and certainly many clouds of lower density.  
What might this mean for time symmetry? 

(For specificity I shall concentrate on photons which are optical today, 
say around 5000\AA.  This band was chosen because at these wavelengths
we have the luxury of the coincidence of decent observations, relatively
well understood theoretical predictions for the background due to galaxies, 
the absence of other plausible sources for significant contributions,
and a respectable understanding of the intergalactic opacity, including
in particular some confidence that the future light cone is transparent.)

Photons at 5000\AA\ today are at the Lyman limit (912\AA) at 
$z \approx 4.5$, and so are ionizing before that.  At these redshifts the 
bounds on the amount of smoothly distributed neutral hydrogen 
(determined by independent measures such as the Gunn-Peterson 
test \cite{Peebles}) are very low ({\it cf.\ }section \ref{sec:flc}),
presumably because that part of the hydrogen formed at recombination
which had not been swept into forming galaxies was ionized by their
radiation.  Before the galactic engines condensed and heated up, however,
this neutral hydrogen would have been very opaque to ionizing radiation.
Similarly, near $z \sim 4$ Lyman-limit clouds with $\Sigma \gtwid
10^{17} {\rm\,cm}^{-2}$ are opaque at these frequencies.  
The upshot is that most photons from our galaxies which are optical today 
will make it well past the time reverse of the present epoch, likely 
ending up in the (time-reversed) 
L-$\alpha$ forest or in a young (to time reversed observers!) galaxy by 
$\tilde{z} \sim 4$.  (Here $\tilde{z}$ is the epoch corresponding to the 
time-reverse of redshift $z$.)  The very few that survive longer must be
absorbed in the sea of neutral hydrogen between $\tilde{z}=0$ and (their) 
recombination epoch, $\tilde{z} \sim 1000$.

Now, the important point is that on average, galaxies radiate
isotropically into the full $4\pi$ of sky available to them.  The lesson
of the previous paragraph is that most lines of sight from galaxies in
the expanding phase will not encounter a high column density cloud until a 
fairly high time-reversed redshift, $\tilde{z} \sim {\rm few}$, 
at which point many lines of sight probably {\it will} intersect 
one of these proto-galaxies or their more diffuse halos.  
If most photons from our galaxies have not been absorbed 
by this point, {\it this is not consistent with time symmetry}:  the 
rate of emission of (what is today) optical radiation by stars in galaxies 
could not be time symmetric if the light of the entire history of galaxies 
in the expanding phase ends up only on the galaxies of the recollapsing 
phase at high $\tilde{z}$ (due consideration of redshifting effects 
is implied, of course.)  Put another way, time symmetry requires the 
specific energy
density in the backround radiation to be time symmetric.  Thus the
emission rate in the expanding era must equal (what we would call) the
absorption rate in the recontracting phase.  If stars in galaxies were
exclusively both the sources (in the expanding phase) and sinks (as we 
would call them in the recontracting phase) of this radiation, galactic
luminosities in the expanding phase would 
have to track 
the falling rate of absorption due to photon ``collisions" with galaxies.  
This is absurd.  At the present epoch, for example, {\it at all frequencies} 
galaxies would (by time symmetry) have to be absorbing the diffuse EGBR (a 
rate for which the upper limit is determined entirely by geometry 
in the ``black galaxy" approximation) at the same rate as their stars were 
radiating (a rate that, in a time symmetric universe which resembles our own,
one expects to be mostly determined by conventional physics.)\footnote{This 
is illustrated in the appendix with a simplified model.
Related considerations may be used to put detailed constraints on the 
self-consistency of time symmetry, but I do not address that any further 
beyond the appendix.  The essential point has already been made.}
That is, stars would be in radiative 
equilibrium with the sky!  This may be called the ``no Olber's Paradox"
argument against the notion that a single class of localized objects could 
be exclusively responsible for the EGBR in a transparent universe equipped 
with time symmetric boundary conditions.  (It might be
thought that this problem would be solved if the time symmetric boundary
conditions lead galaxies to radiate preferentially in those directions 
in which future galaxies lie.  This is not a viable solution, because as
noted above, only a small fraction of the sky is subtended by future 
galaxies up to high time reversed redshift.  The deviations from
isotropic emission would be dramatic.)  

Thus, the option consistent with time symmetry is that most galactic photons
which are optical today will ultimately be absorbed in the many (by time
symmetry) time-reversed Lyman-$\alpha$ forest clouds or Lyman-limit clouds 
believed to dwell between galaxies, and not in the stars of the time-reversed 
galaxies themselves.\footnote{This may be disappointing.  A nice picture of a 
time symmetric universe might have photons from our galaxies arriving at 
time-reversed galaxies in the recollapsing era, appearing as their emissions.  
Even ignoring the highly detailed correlations between the expanding and 
recollapsing phases this would imply, the scheme  could only work if radiation 
could be removed by galaxies in the recollapsing phase at the same rate it is 
emitted in the expanding.  As noted, for isotropically emitting sources this 
is forbidden by time symmetry of the emission rate and geometry.}  Fortunately 
for the notion of time symmetry this indeed appears to be the case.  Careful 
studies of the opacity associated with Lyman systems \cite{Zuo,Madau},
indicates, within the bounds of our rather limited knowledge, that the light 
cone between $z=4.5$ and $z=0$ is essentially totally opaque to radiation that 
is 5000\AA\ at $z=0$, and that this is due largely to Lyman clouds 
near $z \sim 4$ and in the middle range of observed column densities, 
$\Sigma \sim 10^{16-17} {\rm\,cm}^{-2}$ or so.  (To be honest, it must be
admitted that hard data on just such clouds is very limited \cite{MM,Madau}.)

We have now arrived at a terrible conundrum for the notion of time
symmetry.  Even if one is willing to accept the amazingly detailed
correlations between the expanding and recontracting eras that reconciling a
transparent future light cone with time symmetry requires, and even if
the ``excess" radiation correlated with the galaxies of the recollapsing
era were to be observed, this picture is                   
incompatible with what little is known about the physical properties of 
the Lyman-$\alpha$ forest.   Recalling the minimal prediction above for the 
excess background required by time symmetry, the prediction is that the 
Lyman-$\alpha$ forest has produced an amount of radiation at least comparable 
to that produced by the galaxies to our past.  There is no mechanism by which 
this is reasonable.  {\it There is no energy source to provide this amount
of radiation.}  More prosaically, the hydrogen plasma in which the clouds 
largely consist is observed (via determination of the line shape,
for example) to be at kinetic temperatures of order $10^{4-5}K$, heated by
quasars and young galaxies \cite{Carswell,Giallongo}.  Thermal bremsstrahlung 
is notoriously inefficient, and line radiation at these temperatures is 
certainly insufficient to compete with nuclear star burning in galaxies!
At for instance 5000\AA\
today, essentially {\it no} radiation is expected from forest clouds at all, 
let alone an amount comparable to that generated by galaxies.  Remembering
that by redshifts of 4.5 the Lyman forest is essentially totally opaque
shortward of 5000\AA\ (observed) \cite{Madau}, it might have been imagined 
that 
an early generation of galaxies veiled by the forest heated up the clouds
sufficiently for them to re-radiate the isotropic background radiation 
required by time symmetry.  While it is true that quasars and such are
likely sources of heat for these clouds \cite[for example]{Giallongo}, 
aside from the considerable difficulties in getting the re-radiated spectrum 
to resemble that of galaxies, the observed temperatures of the clouds are 
entirely too low to be compatible with the {\it minimum} amount of energy 
emission in the bands required.  

(A related restriction arises from present day observations of cosmological 
metallicities, which constrain the amount of star burning allowed to our past.
If observed discrete sources came close to accounting for the required 
quantity of heavy elements, the contribution of a class of objects veiled 
completely by the Lyman forest would be 
constrained irrespective of observations of 
the EGBR.  However, at present direct galaxy counts only provide 
about 10\% of 
the current upper limits on the extragalactic background light \cite{Tyson}
({\it cf.\ }section \ref{sec:observations}), the
rest conventionally thought to arise in unresolvable galactic sources. 
Consequently, correlating formation of the heavy elements with
observed discrete sources does not at present provide a good test of
time symmetry.  At any rate, such a test is likely to be a
less definitive constraint on time symmetry because it is possible that a 
portion of the radiation lighting the Lyman forest from
high redshift is not due to star burning, but to accretion onto
supermassive black holes at the centers of primordial galaxies.  Thus
the best observational test is the most direct one, comparison of the
observed EGBR with the contribution expected from galaxies.)

The possibility that somehow the excess radiation does {\it not} come 
from the Lyman-$\alpha$ forest, but somehow shines through from other
isotropically distributed sources even further in the past, is 
hardly more appealing.   Familiar physics
tells us that the forest is totally opaque to radiation that is 5000\AA\
at $z=0$.  The conclusion had better be that the universe is not time
symmetric, rather than that time symmetry engineers a clear path only
for those photons correlated with galaxies in the recollapsing epoch
(and not, say, the light from quasars.)  
Moreover, even if that were the case, analagous difficulties apply
to the vast sea of neutral hydrogen that existed after recombination,
totally opaque to ionizing radiation, and again to the highly opaque
plasma which constituted the universe {\it before} recombination. 
It is possible to conjure progressively more exotic scenarios which
save time symmetry by placing the onus on very special boundary conditions 
which engineer such rescues, but this is not the way to do physics.
The only {\it reasonable} way time symmetry could be rescued would be if
it were discovered that for reasons unanticipated here, the future light
cone were not transparent after all, thus obviating the need for an
excess background radiation with all its attendant difficulties.
Otherwise, it is more reasonable to conclude that a universe with time
symmetric boundary conditions would not resemble the one in which we
actually live.

\subsubsection*{Beyond the Minimal Prediction}

Now that we have seen what kind of trouble time symmetry can get into
with only the {\it minimal} required excess background radiation, it is
time to make the problems worse.  The background radiation
correlated with the galaxies of the recollapsing era was bounded from
below, via time symmetry, by including only the radiation that has been
emitted by the galaxies to our past.  But as our stars continue to burn,
if the future light cone is indeed transparent it is possible a great
deal more radiation will survive into the recollapsing era \cite{DT,TSA}. 
How much more?  To get an idea of what's possible it is necessary to
know both what fraction of the baryons left in galaxies will be eventually 
be burned into starlight, and when.  For a rough upper bound, assume that 
 {\it all} of the matter in galaxies today, including the apparently
substantial dark halos (determined by dynamical methods to contribute 
roughly 
$\Omega_{{\rm gal}} \sim .1 $), will eventually be burned into radiation.  
To get a rough lower bound, assume that only the observed luminous matter 
($\Omega_{{\rm lum}} \sim .004$) will participate significantly, and that
only a characteristic fraction of about 4\% of {\it that} will not end up 
in remnants (Jupiters, neutron stars, white dwarfs, brown dwarfs, 
black holes, {\it etc.})  
To overestimate the energy density of this background at 
$\tilde{z} = 0$,
assume that all of this energy is released in a sudden burst 
at some redshift 
$z_{e} (< 0)$.  Then by time symmetry, further star burning will yield a 
background of radiation correlated with time-reversed galaxies (expressed 
as a fraction of the critical density and scaled to $z=0$) 
somewhere in the range
$$ (1+z_{e})^{-1} 10^{-6} \ltwid \Omega_{{\rm burn}}
                          \ltwid (1+z_{e})^{-1} 10^{-3}  \label{Oburn}.$$
(Here I have used the fact that the mass fraction released in nuclear 
burning as electromagnetic radiation is .007.)  
When $(1+z_{e})^{-1} \sim 1$ the upper limit is two orders of magnitude 
more than is in the CMBR today and three orders of magnitude more than 
present observational upper limits on a 
diffuse optical extragalactic background 
({\it cf.\ }section \ref{sec:observations}).   The lower bound,
however, is comparable to the amount of radiation that has already been
emitted by galaxies.  Thus if the lower bound obtains, the prediction for
the optical EGBR in a time symmetric universe is only of order three times
that due to the galaxies to our past (if the excess background inferred
from continued star burning is not distributed over many decades 
in frequency,
and if most of this burning occurs near $z(\tilde{z})=0$.)  
As will be seen in section \ref{sec:observations}, 
this may still be consistent with present
observational upper limits.  On the other hand, if something closer to the 
upper limit obtains this is a clear death blow to time symmetry.
A more precise prediction is clearly of interest.  This would entail
acquiring a detailed understanding of further galactic evolution, 
integrating over future emissions with due attention to the epoch 
at which radiation of a 
given frequency is emitted.  (Naturally, this is the same exercise one 
performs in estimating the EGBR due to galaxies to our past \cite{Deep}.)  

Some idea of the possible blueshift ($(1+z_{e})^{-1}$) involved comes from 
estimating how long it will take our galaxies to burn out.  This should 
not be more than a factor of a few greater than the lifetime of the 
longest lived stars, so a reasonable ballpark figure is to assume that 
galaxies will live for
only another ten billion years or so.   For convenience,  assume that
galaxies will become dark by $ t = n t_{0}$ for some $n$, where $t_{0}$ 
is the 
present age of the universe.  To overestimate the blueshift at this time, 
assume the universe is flat, so that 
$$ (1+z_{e})^{-1} = (t/t_{0})^{\frac{2}{3}} = n^{\frac{2}{3}} .$$
For reasonable $n$'s this does not amount to a large (in order of magnitude)
transfer of energy to the radiation from cosmological recontraction.

\subsubsection*{Additional Sources of ``Excess" EGBR and the Far Future}

In a similar fashion to continued burning of our stars, any isotropic
background produced to our future might by time symmetry be expected 
to imply 
an additional contribution to the EGBR in an appropriately blueshifted
band.  For instance, even if continued star burning does not 
(by time symmetry)
yield a background in contradiction with observations of the EGBR,
it is possible that accretion onto the supermassive black holes likely
to form at the centers of many galaxies could ultimately yield a
quantity of radiation dramatically in excess of that from star burning
alone.\footnote{I owe this suggestion to R. Antonucci.}  
In the absence of detailed information about such possibilities it is
perhaps sufficient to note that ignoring possible additional contributions
leads to a lower limit on the EGBR correlated with sources in the
recontracting era, and I will therefore not consider them.

There is one worrying aspect, however.  As discussed in some detail by
Page and McKee \cite{PM} for an approximately $k=0$ universe, and commented
on in a related context in section \ref{sec:collapse}, if baryons 
decay then considerable photons 
may be produced by for instance the pair annihilation of the resulting 
electrons and positrons.  Should not this, by time symmetry, yield a further 
contribution to the EGBR?  The answer may well be yes, but there is a
possible mechanism which avoids this conclusion.  Somehow, with CPT symmetric 
boundary conditions, the density of baryons must be CPT symmetric.  Therefore 
either baryons do not decay, or they are re-created\footnote{Note that in the 
former case CPT-symmetry requires that the observed 
matter-antimatter asymmetry inside the horizon does not persist 
at larger scales.  In the latter case, {\it if} matter dominates 
homogeneously in the expanding era, then antimatter must dominate 
homogeneously in the recollapsing phase.  Further discussion of 
CP-violation in T- and CPT-symmetric universes may be found in \cite{TSA}.} 
in precisely correlated collisions.  (In the absence of a final boundary
condition, the interaction rate would be too low for (anti-)baryon
recombination to occur naturally.)  The latter (boundary condition
enforced) possibility appears extraordinary, but if baryon decay occurs 
in a universe with CPT symmetric boundary conditions, 
it could be argued that the best electrons and 
photons for the job would be just those created during baryon decay in the 
expanding phase, thereby removing this photon background.  The ``no
Olber's Paradox" argument, that most of an isotropically emitting source's 
light must end up in some homogeneous medium, and not, if time symmetry 
is to be preserved, equivalent time-reversed point sources, may not apply 
here if matter is relatively homogeneously distributed when the universe 
is large.  
Baryon decay {\it might} smooth out 
inhomogeneities somewhat before the resulting electrons and positrons 
annihilate.  (This requires the kinetic energies of the decay products to be 
comparable to the gravitational binding energy of the relevant inhomogeneity.)
Then the picture is no longer necessarily of localized sources emitting 
into $4\pi$, but of a more homogeneous photon-producing background 
that might cover 
enough of the sky to more reasonably secrete the required correlations for 
reconstruction of more massive particles in the recontracting phase.  
Nevertheless the extreme awkwardness of this scenario is not encouraging.
The former possibility, clearly more palatable, is that baryons do not
decay significantly either because $\Omega$ is not so near one after all 
that they have time enough to do so, or because the presence of the final
boundary condition suppresses it.  Either way, in this (possibly desperate)
picture there is no additional background due to decaying baryons.
A very similar question relates to the enormous number of particles
produced in the last stages of black hole evaporation.  This
time, however, the objection that our black holes cover only a small
portion of the recontracting era's sky, and consequently their isotropic
emissions could not do the job of forming the white holes of the 
recontracting
era (black holes to observers there) time symmetrically, 
would seem to be forceful.  Thus if the universe is indeed very
long-lived, black hole evaporation may well require yet an additional 
observable background.  This may not be such a serious difficulty if
$\Omega $ is not very close to one, however, as the time scales for the 
evaporation of galactic-scale black holes are quite immense.  Further 
discussion of black holes in time symmetric cosmologies may be had in 
\cite{Zeh3,Penrose1,Penrose2}.

One last point regarding the predictions described in this section needs
to be made.  Clearly, a loose end which could dramatically change the 
conclusions is the condition of matter in the universe 
when it is very large.  
This is uncertain territory, not the least because that is the era in a 
statistically time symmetric universe when the thermodynamic arrow 
must begin rolling over.  Neglecting this confusing complication 
(reasonable for some purposes as many interactions are most significant 
when the universe is small), there is not a great deal known about 
what the far future should look 
like \cite{Rees,Davies,Islam,BT,Dyson,PM}.  The study of Page and 
McKee \cite{PM} gives the most detailed picture in the case of a 
flat universe.
As mentioned at the end of section \ref{sec:flc}, Davies and Twamley 
\cite{DT} find from this 
work that interactions of optical (at $z=0$) backgrounds with the electrons 
produced by baryon decay do not appear to be significant, primarily due to 
their diffuseness.  On the other hand, if supermassive black holes (or any 
large gravitational inhomogeneities) appear, 
interactions with them may induce 
anisotropies in the future starlight.  However, clumping only decreases 
the probability a line of sight intersects a matter distribution.  
Therefore large overdensities probably never subtend enough 
solid angle to interfere with most lines of sight to the recollapsing 
era unless gravitational collapse proceeds
to the point where it dramatically alters homogeneity and isotropy on
the largest scales.  Because collapse is rather strongly constrained
by time symmetry ({\it cf.\ }section \ref{sec:collapse}) I will not 
consider this possibility.  
Thus, insofar as the prediction of an ``excess" background radiation 
correlated with galaxies in the recollapsing era is concerned, 
the state of the universe when it is large would does not obviously 
play a substantial role.  Nevertheless, given the manifold difficulties 
cited, the sentiment expressed 
above is that the best hope a time symmetric model has of providing a 
realistic description of our universe is that some unforseen mechanism 
makes the future light cone opaque after all.

\subsection*{Summary}     
\label{sec:opacitysummary}

To summarize, because our future light cone is transparent, neglecting
starburning to our future and considering only the contribution to the 
EGBR 
from stars in our past provides an estimate of the total EGBR correlated
with galaxies in the recollapsing era that is actually a lower limit on it.
As mechanisms for distorting the spectrum generically become less important 
as the universe expands (barring unforseen effects in the far future), it is 
reasonable to take models for the present EGBR due to stars in our past as a 
minimal estimate of the isotropic background of starlight that will make its 
way to the recollapsing era.  By time symmetry we can expect that at the same 
scale factor in the recollapsing era similar (but time-reversed) conditions 
obtain.  As argued above, by time symmetry and geometry this 
``future starlight" must appear to us as an 
additional background emanating from homogeneously distributed sources 
to our past {\it other} than galaxies.  Therefore, if the universe has 
time symmetric 
boundary conditions which (more or less) reproduce familiar physics when the 
universe is small, and our future light cone is transparent, the optical 
extragalactic background radiation should be at least twice that expected to 
be due to stars in our past alone, and possesses a similar spectrum.  If a 
considerable portion of the matter presently in galaxies will be burned into 
radiation in our future, by time symmetry the expected background is 
potentially much larger, and observations of the EGBR may already be flatly 
incompatible with observations.  Nevertheless, in the
next section I shall be conservative and stick with the minimal prediction
in order to see how it jibes with observations.

\subsection{Models and Observations of the Optical EGBR}
\label{sec:observations}

At optical wavelengths, it is generally believed that the isotropic
background of radiation from extragalactic sources is due entirely to
the galaxies on our past light cone \cite{Deep,IAU,EGBR}. As shown 
in the previous section, if our universe is time symmetric there must in 
addition be a significant contribution correlated with the galaxies of the
recollapsing era which arises, not in galaxies, but in some homogeneously 
distributed medium, say for instance the Lyman clouds.
The apparent inconsistency of this prediction with what is known about
the forest clouds has been discussed above,
and may be taken as an argument that our universe is not time symmetric.
In this section judgement will be suspended, and the prediction of an
``excess" EGBR at least comparable to, but over and above, that due to
galaxies to our past will be compared with experiment.   The conclusion
will be that current data are still consistent with time symmetry {\it if}
our galaxies will not, in the time left before they die, emit a quantity 
of radiation that is considerably greater than that which they already have.

A useful resource on both the topics of this section is \cite{Deep}.

Tyson \cite{Tyson} has computed how much of the optical extragalactic
background is accounted for by resolvable galaxies, concluding that
known discrete sources contribute
\begin{equation}
\label{eq:tyson}
\nu i_{\nu} \sim 3 \cdot 10^{-6} {\rm\,erg}\, {\rm\,cm}^{-2}\, {\rm\,s}^{-1}\, 
 {\rm\,ster}^{-1}
\end{equation}
at 4500\AA.  However, because very distant galaxies contribute most of
the background radiation it is believed that unresolvable sources provide
a significant portion of the EGBR.  At present it is not possible to
directly identify this radiation as galactic in origin.  However, as
understanding of galactic evolution grows so does the ability to model
the optical extragalactic background due to galaxies.  These predictions
naturally depend on the adopted evolutionary models, what classes of
objects are considered, the cosmological model, and so on.  As
representative samples I quote the results of Code and Welch \cite{CW}
for a flat universe in which all galaxies evolve,
\begin{equation}
\label{eq:CW}
\nu i_{\nu}\sim 8\cdot 10^{-6} {\rm\,erg}\, {\rm\,cm}^{-2}\, {\rm\,s}^{-1}\,
 {\rm\,ster}^{-1}
\end{equation}
at 5000\AA, and of Cole {\it et al.\ }\cite{CTS} for a similar scenario,
\begin{equation}
\label{eq:Cole}
\nu i_{\nu}\sim 3\cdot 10^{-6} {\rm\,erg}\, {\rm\,cm}^{-2}\, {\rm\,s}^{-1}\,
 {\rm\,ster}^{-1},
\end{equation}
also at 5000\AA.  These figures are to be compared with the results of
(extraordinarily difficult) observations.   As surveyed by
Mattila \cite{Mattila}, they give at 5000\AA\ an upper limit of
\begin{equation}
\label{eq:mattila}
\nu i_{\nu}\ltwid 2\cdot 10^{-5}{\rm\,erg}\, {\rm\,cm}^{-2}\, {\rm\,s}^{-1}\,
 {\rm\,ster}^{-1}.
\end{equation}
As far as I am aware, there has been no direct detection of an optical
radiation background of extragalactic origin.  This upper limit represents 
what is left after what can be accounted for in local sources is removed.

Comparing these results, it is clear that if current models of galactic
evolution are reliable, present observations of the extragalactic background 
radiation leave room for a contribution from non-galactic sources that
is comparable to the galactic contribution, but not a great deal more.
These observations therefore constrain the possibility that our
universe is time symmetric.  If believable models indicate that further
galactic emissions compete with what has been emitted so far, time
symmetry could already be incompatible with experiment.  A direct detection 
of the extragalactic background radiation, or even just a better upper limit, 
could rule out time symmetry on {\it experimental} grounds 
soon.\footnote{To be more careful, if direct HST galaxy counts,
or reliable models of the (extra-)galactic
contribution closely agree with the observations, the justifiable
conclusion is that {\it if} our universe is time symmetric, then for some
unknown reason our future light cone is not transparent.}  (The ideal
observational situation would be convergence of all-sky photometry and
direct HST galaxy counts, allowing one to dispense with models
completely.)

\section{Further Difficulties with Time Symmetry}
\label{sec:difficulties}
\setcounter{equation}{0}

In this section I comment on some issues of a more theoretical nature which 
must be faced in any attempt to construct a believable model of a time 
symmetric universe.  Among other questions, in such a universe the 
difficulties with incorporating realistic gravitational collapse, and in 
deriving the fact that radiation is retarded, appear to be considerable.

\subsection{Gravitational Collapse}
\label{sec:collapse}

A careful account of the growth of gravitational inhomogeneities from the 
very smooth conditions when the universe was small is clearly of fundamental 
importance in any model of the universe, time symmetric or not, not the least 
because it appears to be the essential origin of the thermodynamic arrow of 
time\footnote{For examples of concrete calculations connecting the growth 
of gravitational inhomogeneities with the emergence of a thermodynamic arrow 
see \cite{HaH,Hawking1,HLL}.}
from which the other apparent time arrows are thought to flow \cite{PTA,Zeh2}.
However, if matter in a closed universe is to be smooth at 
both epochs of small
scale factor then it is incumbent to demonstrate that Einstein's equations 
admit solutions in which an initially smooth universe can develop interesting 
(non-linear) inhomogeneities such as galaxies which eventually 
smooth out again 
as the universe recollapses.  This is because the universe is certainly in
a quasiclassical domain now, and if it is assumed to remain so
whenever the scale factor is large classical general relativity must
apply.  Laflamme \cite{Laflamme} has shown that in the linear regime there 
are
essentially no perturbations which are regular, small, and growing away
from both ends of a closed FRW universe, so that in order to be small at 
both
ends a linear perturbation must be too small to ever become non-linear.
But that is not really the point.  Interesting perturbations must go
non-linear, and there is still no proof of which I am aware that
perturbations which go non-linear cannot be matched in the non-linear
regime so as to allow solutions which are small near both singularities.
Put differently, what is required is something like a proof that 
Weyl curvature 
must increase ({\it cf.\ }\cite{Penrose1,Penrose2}), {\it i.e.\ }that given 
some suitable energy conditions the evolution of gravitational inhomogeneity 
must be monotonic even in the absence of trapped surfaces, 
except possibly on a set of initial data of measure zero.  
While perhaps plausible given the attractive nature of gravity, proof has 
not been forthcoming.

(It is important for present purposes that genericity conditions 
on the initial data are not a central part of such a proof, for physically
realistic solutions which meet the time symmetric boundary conditions
must describe processes in the recollapsing era which from our point of
view would look like galaxies disassembling themselves.  Reducing such a
solution to data on one spacelike hypersurface at, say, maximum
expansion, said data will be highly specialized relative to solutions
with galaxies which do not behave so unfamiliarly.  If such solutions
with physically interesting inhomogeneities do exist, the real question
here is whether they are generic {\it according to the measure defined
by the two-time boundary conditions.}  Since it ought to be possible to 
treat this problem classically, in principle this measure is straightforward 
to construct.  The practical difficulty arises in evaluating the generic 
behaviour of solutions once they go non-linear.  My own view is that it is 
highly likely that such solutions remain exceedingly improbable according to 
a measure defined by a {\it generic} set of (statistical) boundary conditions 
requiring the universe to be smooth when small.  As noted, Laflamme 
\cite{Laflamme} has already shown that when the initial and final 
states are required to be smooth,
the growth of inhomogeneity is suppressed if only linear perturbations 
are considered.  Unless boundary conditions with very special correlations 
built in are imposed, probable solutions should resemble their smooth initial 
and final states throughout the course of their evolution, never developing 
physically interesting inhomogeneities.\footnote{One possible out 
is Schulman's observation that systems which exhibit chaos, as
general relativity does, may be less restricted in the varieties of
their behaviour by boundary conditions at two times than are systems 
with linear dynamics \cite{Schulman3}.  This substantially unstudied
possibility would obviously never emerge from Laflamme's linear 
analysis.}

Note the concern here is not with occurrences which would be deemed 
unlikely in a universe with an initial condition only, but occur in a 
time symmetric universe because of the ``fate" represented by the final 
boundary condition, but with occurrences which are unlikely {\it even in a 
universe with (generic) time symmetric boundary conditions.})

A related question concerns collapsed objects in a time
symmetric universe.  Page and McKee \cite{PM}
have studied the far future of a $k=0$ FRW universe under the
assumption that baryons decay but electrons are stable.  Assuming
insensitivity to a final boundary condition their conclusions should
have some relevance to the period before maximum expansion in the only
slightly over-closed (and hence very long-lived) model universes that
have mostly been considered here.  As discussed in section
\ref{sec:amplifications}, if the universe is very long lived it might be
imagined that the decay of baryons and subsequent annihilation of the
produced electrons and positrons could smooth out inhomogeneities, and
also tend to destroy detailed information about the gravitational history of 
the expanding phase (by eliminating compact objects such as neutron stars,
for instance.)  Thus, even though interactions are unlikely to thermalize 
matter and radiation when the universe is very large \cite{PM} 
({\it cf.\ }section \ref{sec:opacity}), there may be an analagous 
information loss via 
the quantum decay of baryons which could serve a similar function.  
(For a completely different idea about why quantum mechanics may effectively 
decouple the expanding and recollapsing eras, see \cite{Zeh1,Zeh2,Zeh3,KZ}.)
In any case, if there is no mechanism to eliminate collapsed objects
before the time of maximum expansion, then the collapsed objects of the
expanding phase are the same as the collapsed objects of the
recontracting phase, implying detailed correlations between the
expanding and recontracting era histories of these objects which might
lead to difficulties of consistency with the RTH.  

A particular complication is that it is fairly certain that black
holes exist, and that more will form as inhomogeneity grows.  The only
way to eliminate a black hole is to allow it to evaporate, yet the
estimates of Page and Mckee indicate that it is more likely for black
holes to coalesce into ever bigger holes unless for some reason (a final
boundary condition?) there is a maximum mass to the black holes which
form, in which case they may have time enough to evaporate (though this
requires $\Omega$ to be {\it exceedingly} close to one.)
In fact, it may be imperative for a time symmetric scenario that black holes 
evaporate, else somehow they would have to turn into the white holes of the
recollapsing era (black holes to observers there.)  This is because we do 
not observe white holes today \cite{Penrose1,Penrose2}.  (A related
observation is that in order for the universe to be smooth whenever it
is small, black/white hole singularities cannot arise \cite{Zeh3}.)
Here the evaporation of black holes before
maximum expansion would be enforced by the time symmetric boundary
conditions selecting out only those histories for which there are no
white holes in the expanding era and {\it mutatis mutandis} for the
recollapsing era.\footnote{For more on black holes in time symmetric 
universes, see the discussion of Zeh \cite{Zeh3}.}  
On the other hand, if evaporating black holes leave remnants they too 
must be worked into the picture.  Again, if the results of 
the stochastic models with two-time low entropy boundary conditions 
discussed above are to be taken seriously, the conclusion should probably 
be that boundary conditions requiring homogeneity when the universe is 
small suppress histories in which significant gravitational collapse 
occurs by assigning low probabilities to histories with fluctuations 
that will go non-linear.  It hardly needs emphasizing that all of these 
considerations are tentative, and greater clarity would be welcome.


\subsection{The Retardation of Radiation}
\label{sec:retardation}

Besides gravitational considerations, radiation which connects the
expanding and recollapsing eras provides another example of a physical
process which samples conditions near both boundary conditions \cite{DT}.
While gravitational radiation and neutrinos are highly penetrating and are
likely to provide such a bridge, in neither case are we yet capable of both 
effectively observing, and accurately predicting, what is expected
from sources to our past.  Therefore the 
primary focus of this investigation has been electromagnetic radiation.
Above it was confirmed that modulo the obviously substantial uncertainty
regarding the condition of the universe when it is large, even
electromagnetic radiation is likely to penetrate to the recollapsing
era.  Section \ref{sec:egbr}\ was concerned with one relatively prosaic 
consequence of this prediction if our universe possesses time symmetric 
boundary conditions.  Here I comment briefly on another.

Maxwell's equations, the dynamical laws governing electromagnetic
radiation, are time symmetric.  It is generally believed that the
manifest asymmetry in time of radiation {\it phenomena}, that is, that 
(in the absence of source-free fields) observations are described by 
retarded solutions rather than advanced, is ascribable fundamentally 
to the thermodynamic arrow of time without additional hypotheses.  
(For a contemporary review see \cite{Zeh2}.)
However, if our universe possesses time symmetric boundary conditions
then near the big bang the thermodynamic arrow of entropy increase runs
oppositely to that near the big crunch.  Since radiation can connect the
expanding and recollapsing eras, the past light cone of an accelerating
charge in the expanding era ends up in matter for which the entropy is
increasing, while its future light cone terminates in matter for which
entropy is supposed to be {\it decreasing}.  If the charge radiates into
its future light cone this implies detailed correlations in this matter
with the motion of the charge which are incompatible with the supposed
entropy decrease there (entropy increase to time reversed observers),
although it is true that these correlations are causally disconnected to
time-reversed observers, and consequently invisible to {\it local}
coarse grainings defining a local notion of entropy for such observers.
This state of affairs makes it difficult to decide whether radiation
from an accelerating charge (if its radiation can escape into intergalactic 
space) should be retarded 
(from the perspective of observers in the expanding era) 
or advanced or some mixture of the two. The conditions under which the 
radiation arrow is usually derived from the 
thermodynamic arrow of surrounding 
matter do not hold.  (Notice how this situation is reminiscent of the 
requirements necessary to derive retardation of radiation in the 
Wheeler-Feynman ``absorber theory" of electrodynamic phenomena \cite{WF}.)
Hence the ability of radiation to connect the expanding and recollapsing
epochs brings into question the self-consistency of assuming
time symmetric boundary conditions on our universe together with 
``physics as usual" (here meaning radiation which would be described as
retarded by observers in both the expanding and recollapsing eras) near 
either end.  The retardation of radiation is another important example 
of a physical prediction which would be expected to be
very different in a universe with time symmetric boundary conditions
than in one without.  Once again, if the results of the simple stochastic 
models are generally applicable, the retardation of radiation should 
no longer be a prediction in such a universe.\footnote{Analytical 
elucidation 
of this idea is hoped to be the subject of a (far) future paper.}  

To summarize this section, consideration of gravitational collapse and 
radiation phenomena reveals that construction of a model universe with
time symmetric boundary conditions which resembles our own may be a 
difficult task indeed.  There are strong suggestions that a 
model with time symmetric boundary conditions which mimic our own
early universe would behave nothing like the universe in which we live.
Such a model would most likely predict a universe which remained smooth
throughout the course of its evolution, with coupled matter components
consequently remaining in the quasi-static ``equilibrium" appropriate to a
dynamic universe.

\section{Summation}
\label{sec:summation}
\setcounter{equation}{0}

In spite of the oft-expressed intuitive misgivings regarding the
possibility that our universe might be time symmetric \cite[for example]
 {Weinberg,PTA,Penrose1}, it has generally been felt that if sufficiently 
long-lived, there might be no way to tell the difference between a time 
symmetric and time asymmetric universe.  Building on suggestions of
Cocke, Schulman, Davies, and Gell-Mann and Hartle (among others), this
work has explored in some detail one physical process which, happily,
belies this feeling:  no matter how long our universe will live, the
time symmetry of the universe implies that the extragalactic background
radiation be {\it at least} twice that due to the galaxies to our past.
This is essentially due to the fact that light can propagate unabsorbed
from the present epoch all the way to the recollapsing era.
Moreover, geometry and time symmetry requires this ``excess" EGBR to be
associated with sources {\it other} than the stars in those galaxies, 
sources which, according to present knowledge about the era during which
galaxies formed, are not capable of producing this radiation!  Thus the
time symmetry of a closed universe is a property which is {\it directly
accessible to experiment} (present observations are nearly capable of
performing this test), as well as extremely difficult to model convincingly 
on the basis of known astrophysics.  In addition, the other theoretical 
obstacles remarked upon briefly in sections \ref{sec:amplifications}\
and \ref{sec:difficulties}\ make it difficult to see how a 
plausible time symmetric model for the observed universe 
might be constructed.
In particular, any such attempt must demonstrate that in a universe that is 
smooth whenever it is small, gravitational collapse can procede to an 
interesting degree of inhomogeneity when the universe is larger 
\cite{Laflamme}.
This is necessary in order that the universe display a thermodynamic arrow 
(consistently defined across spacelike slices)
which naturally reverses itself as the universe begins to recollapse.
Furthermore, it appears unlikely that the usual derivation of the
retardation of radiation will follow through in a time symmetric
universe in which radiation can connect regions displaying opposed
thermodynamic arrows.  Finally, in the context of the time neutral
generalized quantum mechanics employed as the framework for this
discussion, unless the locally observed matter-antimatter asymmetry
extends globally across the present universe,\footnote{Recall that
this requires that the universe live long enough for nearly all baryons
to decay, and reform into the antibaryons of the recollapsing era.  
This presents serious additional difficulties, 
 {\it cf.\ }section \ref{sec:amplifications}.}
natural choices of CPT-related boundary conditions yield a theory with 
trivial dynamics if the deviations from exact homogeneity and isotropy 
are specified in a CPT invariant fashion (see the last paragraph of section
\ref{sec:motivations}). 
In sum, were the ``excess" EGBR which has been the primary concern of this 
investigation to be observed, it would appear necessary to place the onus of 
explanation of the fact that the final boundary condition is otherwise
practically invisible upon very specially chosen boundary conditions which
encode the details of physics in our universe.   This would make it difficult 
to understand these boundary conditions in a natural way.  
On the dual grounds 
of theory and experiment, it therefore appears unlikely that we live in a 
time symmetric universe.  (A definitive expurgation must await more thorough 
investigation of at least some of the aforementioned difficulties.)

\appendix

\section*{Appendix: No Time Symmetric Olber's Paradox}
\label{sec:noolbers}
\setcounter{equation}{0}
\renewcommand{\thesection}{A}

For the malcontents in the audience, this appendix offers a flat space
model explicitly illustrating the ``no Olber's Paradox" argument of 
section \ref{sec:amplifications}.  As cosmological redshifting is time 
symmetric, the complications due to curvature are inconsequential for
present purposes.  (Curvature may be included in a straightforward 
fashion, but that and many other embellishments are, out of courtesy,
foregone.)  Therefore, consider the universe of figure 1
as flat.  For convenience, relocate the zero of conformal time $ \eta $ to
be at the moment of maximum expansion.  The specific energy density in
radiation obeys a transfer equation
\begin{equation}
\label{eq:transfer}
   \frac{d\epsilon }{d\eta } = j - \Sigma \epsilon .
\end{equation}
Here $ j $ represents sources (according to expanders; thus in the
recontracting era $ j $ may be negative), and $ \Sigma $ sinks (same
comment), of radiation.  Time symmetry implies that
\begin{equation}
\label{eq:TSenergy}
    \epsilon (\eta ) = \epsilon (-\eta ),
\end{equation}
\begin{equation}
\label{eq:TSsource}
    j(\eta ) = - j(-\eta ),
\end{equation}
and
\begin{equation}
\label{eq:TSsink}
    \Sigma (\eta ) = - \Sigma (-\eta ).
\end{equation}

Now, suppose it is imagined that a time symmetric universe contains only one 
class of localized, homogeneously and isotropically distributed sources 
({\it i.e.\ }galaxies) 
in the expanding era, with corresponding time-reversed sinks (in the
language used by expanding era observers) in the recontracting era ({\it
i.e.\ }thermodynamically reversed galaxies.)  For isotropically emitting
sources, in the ``black galaxy" approximation the absorption rate
(emission rate to thermodynamically reversed observers) in the
recontracting era can be thought of as being controlled by the amount of
radiation from the expanding era which the recontracting era's galaxies
intercept. That is, the thermodynamically reversed observers of the
recontracting era would see their galaxies emitting at a rate given 
(at most) by
\begin{eqnarray}
\label{eq:j+}
  j(\eta ) & = & - \Sigma (-\eta )\epsilon (-\eta ) \nonumber \\
            & = &  \Sigma (\eta )\epsilon (\eta ),
\end{eqnarray}
using time symmetry, and where $ \left|\Sigma\right| $
is as in equation (\ref{eq:hardspheresigma}).
(This is merely the expression of the fact that in the essentially 
geometric ``black galaxy" approximation, galaxies do not care if they
are intercepting radiation ``from" the past or the future.)  But from
this it is obvious that
\begin{equation}
\label{eq:olbers}
   \frac{d\epsilon }{d\eta } = 0,
\end{equation}
and galaxies are in radiative equilibrium with the sky, a situation
reminiscent of the historically important Olber's Paradox (``Why is the
night sky dark?")  Thus in a transparent, time symmetric universe in
which the night sky is dark, there must be an additional class of 
sources emitting the radiation which is correlated with the galaxies 
of the recontracting era.  

(In reality, $j_{{\rm gal}} \gg \Sigma_{{\rm gal}}\, \epsilon $.  Indeed,
\begin{eqnarray}
\frac{j_{{\rm gal}}}{\Sigma_{{\rm gal}} c\epsilon} 
               & = & \frac{n L_{*}}{nc\sigma\epsilon} \nonumber \\
               & = & \frac{L_{*}}{4\pi\sigma\nu i_{\nu }} \nonumber \\
               & \sim & 10^2,
\end{eqnarray}
using a characteristic galactic luminosity 
$L_{*} = 3.9 \cdot 10^{43}\, {\rm erg\, s^{-1}}$ \cite{Peebles}, 
$\sigma \sim 3 \cdot 10^{45}\, {\rm cm^{-2}}$
({\it cf.\ }(\ref{eq:HS})), (\ref{eq:CW}), and taking $h = 1$.  
It should not come as a surprise that the energy density in the 
optical EGBR is still increasing!)

\section*{Acknowledgements}

  I wish to thank R.\ Antonucci, O.\ Blaes, T.\ Hurt, and R.\ Geller 
for useful conversations on matters astrophysical, J.~T.\ Whelan 
for discussions, comments, and questions, and J.~B.~Hartle for raising 
this question and for provocative discussions.  This work was supported 
in part by NSF Grant No. PHY90-08502.


\end{document}